%% file: Star25MAY2016.tex
\documentclass[twocolumn,traditabstract,longauth,a4]{aa}

\input Planck.tex
\usepackage{graphicx,amsmath}
\usepackage{epsf}
\usepackage{txfonts}
\usepackage[hyphens]{url}
\usepackage{longtable}
\usepackage{lscape}
\usepackage[hyperindex,breaklinks=true, colorlinks, citecolor=blue]{hyperref}
\usepackage{breakurl}
\usepackage{comment}
\usepackage{natbib}
\usepackage{upgreek}
\usepackage{float}
\usepackage[switch,pagewise]{lineno}
\usepackage{multirow}
\usepackage{enumerate}

\newcommand{\hpeter}[1]{}
\newcommand{\commentproof}[1]{{\bf \color{green}#1}}
\renewcommand{\commentproof}[1]{} % turn off mark-up

\include{Planck}

\newcommand{\planck}{\Planck}  % for \planck rather than \Planck

\newcommand{\nh}{$N_{\textsc{H}}$}
\newcommand{\nhd}{N_{\textsc{H}}} % use in math mode

\newcommand{\lognh}{$\log_{10}(N_{\textsc{H}}/\mbox{cm}^{-2})$}

\newcommand{\stwo}{$S_{2}(\ell)$}
\newcommand{\stwostar}{$S^{\rm star}_{2}(\ell)$}
\newcommand{\stwosubmm}{$S^{\rm submm}_{2}(\ell)$}
\newcommand{\stwoallsubmm}{$S^{\rm all\,submm}_{2}(\ell)$}

\newcommand{\bvec}{$\vec{B}$}

\newcommand{\bperp}{$\langle\hat{\vec{B}}_{\perp}\rangle$}
\newcommand{\bperpplanck}{$\langle\hat{\vec{B}}^{\rm submm}_{\perp}\rangle$} 
\newcommand{\bperpstars}{$\langle\hat{\vec{B}}^{\rm star}_{\perp}\rangle$}

\newcommand{\IRAS}{\textit{IRAS\/}}

\newcommand{\healpix}{{\sc HEALPix}}

\providecommand{\sorthelp}[1]{}

\begin{document}
%%%%%%%%%%%%%%%%%%%%%%%%%%%%%%%%%%%%%%%%%%%%%%%%%%%%%%%%%%%%
%\linenumbers
%%%%%%%%%%%%%%%%%%%%%%%%%%%%%%%%%%%%%%%%%%%%%%%%%%%%%%%%%%%%

%\input PIP_113_Boulanger_authors_and_institutes.tex

\title{Magnetic field morphology in nearby molecular clouds as revealed by starlight and submillimetre polarization.}
\titlerunning{Magnetic field morphology as revealed by dust.}
    \author{J.~D.~Soler$^{1,2}$\thanks{Corresponding author: juan-diego.soler-pulido@cea.fr} \and F.~Alves$^{3}$ \and F.~Boulanger$^{2}$ \and A.~Bracco$^{1}$ \and E.~Falgarone$^{4,5,6}$ \and G.\,A.\,P.~Franco$^{7}$ \and V.~Guillet$^{2}$ \and P.~Hennebelle$^{1}$ \and F.~Levrier$^{4,5,6}$ \and P.~G.~Martin$^{8}$\and M.-A.~Miville-Desch\^{e}nes$^{2}$}
    \institute{Laboratoire AIM, Paris-Saclay, CEA/IRFU/SAp - CNRS - Universit\'{e} Paris Diderot, 91191, Gif-sur-Yvette Cedex, France 
\and Institut d'Astrophysique Spatiale, CNRS (UMR8617) Universit\'{e} Paris-Sud 11, B\^{a}timent 121, Orsay, France 
\and Max-Planck-Institut für extraterrestrische Physik, Giessenbachstrasse 1, 85748 Garching, Germany 
\and LERMA, Observatoire de Paris, PSL Research University, CNRS, UMR 8112, F-75014, Paris France 
\and Sorbonne Universit\'es, UPMC Univ. Paris 6, UMR 8112, LERMA, F-75005, Paris, France 
\and LRA, ENS Paris, 24 rue Lhomond, 75231 Paris Cedex 05, France
\and Departamento de F\'{i}sica-ICEx-UFMG, Caixa Postal 702, 30.123-970 Belo Horizonte, Brazil
\and CITA, University of Toronto, 60 St. George St., Toronto, ON M5S 3H8, Canada}

\authorrunning{J.~D.~Soler et al.}

\date{Received XX May 2016 / Accepted XX XXX 2016}

\abstract{
Within four nearby ($d < 160$\,pc) molecular clouds, we statistically evaluate the structure of the interstellar magnetic field, projected on the plane of the sky and integrated along the line of sight, as inferred from the polarized thermal emission of Galactic dust observed by \Planck\ at 353\,GHz and from the optical and NIR polarization of background starlight.
We compare the dispersion of the field orientation directly in vicinities with an area equivalent to that subtended by the \planck\ effective beam at 353\,GHz (10\arcmin) and using the second-order structure functions of the field orientation angles.

We find that the average dispersion of the starlight-inferred field orientations within 10\arcmin-diameter vicinities is less than 20\deg, and that at these scales
the mean field orientation is on average within 5\deg\ of that inferred from the submillimetre polarization observations in the considered regions.

We also find that the dispersion of starlight polarization orientations and the polarization fractions within these vicinities are well reproduced by a Gaussian model of the turbulent structure of the magnetic field, in agreement with the findings reported by the \Planck\ collaboration at scales $\ell > 10$\arcmin\ and for comparable column densities.

At scales $\ell > 10$\arcmin, we find differences of up to 14\pdeg7 between the second-order structure functions obtained from starlight and submillimetre polarization observations in the same positions in the plane of the sky, but comparison with a Gaussian model of the turbulent structure of the magnetic field indicates that these differences are small and are consistent with the difference in angular resolution between both techniques.

The differences between the second-order structure functions calculated with each technique suggests that the increase in the angular resolution obtained with the starlight polarization observations does not introduce significant corrections to the dispersion of polarization orientations used in the calculation of the molecular-cloud-scale magnetic field strengths reported in previous studies by the \Planck\ collaboration.
}

\maketitle
%\allearlypapers
%\tableofcontents

\section{Introduction}\label{section:introduction}

Polarization observations -- in extinction from background stars and emission from dust -- reveal the orientation of the interstellar magnetic field averaged along the line of sight (LOS) and projected on the plane of the sky \citep[\bperp,][]{hiltner1949,davis1951,hildebrand1988,planck2014-XIX}. 
These observations constitute a crucial dataset to study the role of the magnetic field in the formation and evolution of molecular clouds (MCs) and their substructures, from filaments to cores and eventually to stars \citep{bergin2007,mckee2007,crutcher2012}.

Recent observations by \Planck\footnote{\Planck\ (\url{http://www.esa.int/Planck}) is a project of the European Space Agency (ESA) with instruments provided by two scientific consortia funded by ESA member states (in particular the lead countries France and Italy), with contributions from NASA (USA) and telescope reflectors provided by a collaboration between ESA and a scientific consortium led and funded by Denmark.} 
\citep{planck2014-a01} have produced the first all-sky map of the polarized emission from dust at submillimetre wavelengths. 
Compared with earlier ground-based and balloon-borne observations, this survey is an immense step forward in sensitivity, coverage, and statistical significance. 

The studies by the \Planck\ collaboration include an overview of polarized thermal emission from Galactic dust \citep{planck2014-XIX}, which reported polarization 
fractions up to 20\,\% at low total gas column density (\nh), decreasing systematically with increasing \nh\ to a low plateau for regions with $\nhd > 10^{22}$\,cm$^{-2}$.

\cite{planck2014-XX} presented a comparison of this polarized thermal emission towards molecular clouds with results from simulations of magnetohydrodynamic (MHD) turbulence, identifying an anti-correlation between the polarization fraction and the dispersion of the polarization angle that can be understood in terms of the turbulent structure of the magnetic field. 

Over most of the sky, \cite{planck2014-XXXII} analysed the relative orientation between density structures and polarization, revealing that most of the elongated structures (filaments or ridges) are predominantly aligned with the magnetic field measured on the structures. This statistical trend becomes less striking for increasing column density. 

At the scales of MCs, \cite{planck2014-XXXIII} studied the polarization properties of three nearby filaments, showing by geometrical modelling that the magnetic field in those representative regions has a well-defined mean direction that is different from the field orientation in the surroundings. At the same scales, \cite{planck2015-XXXV} showed that the relative orientation between the column density structures and \bperp\ in ten nearby ($d < 450$\,pc) MCs is consistent with what can be derived from simulations of trans- or sub-Alfv\'enic magnetohydrodynamic turbulence in molecular clouds.

Given the 10\arcmin\ effective angular resolution of the 353\,GHz polarization observations\footnote{The nominal resolution of the \Planck\ 353\,GHz band is 4\parcm8, but \cite{planck2015-XXXV} convolve these observations with a Gaussian beam to increase the signal-to-noise ratio (SNR) towards the studied regions.}, \Planck\ samples \bperp\ in physical scales down to 0.4\,pc in the nearest MC in the present work \citep[Taurus, at 135\,pc,][]{schlafly2014}.
In contrast, starlight polarization observations provide estimates of \bperp\ down to scales comparable to the angular diameter of stars. 
Although those observations are limited to 
%optically thin 
lines of sight with moderate dust extinction
towards background stars, they provide insight into the structure of the field at scales that are not accessible to \Planck. 
Previous works by the \Planck\ collaboration used starlight polarization observations to study the properties of dust grains \citep{planck2014-XXI}, but there was no study focused on characterizing the \bperp\ structure within the \planck\ beam and towards MCs.

In the present work, we compare the magnetic field orientations inferred from the observations of optical and NIR starlight polarization, \bperpstars, and those derived from the \Planck\ 353\,GHz polarization observations, \bperpplanck, towards four nearby MCs, namely Taurus, Pipe Nebula, Lupus I, and Musca. 
Given the difference in the angular resolution and LOS depth in each technique, we aim to characterize the structure of the field within the \planck\ beam and 
evaluate the contribution of different portions of the LOS to the observed \bperpplanck.
For that purpose, we evaluate the dispersion of \bperp\ orientations within 10\arcmin\ vicinities and compare the second-order structure functions of \bperpstars\ and \bperpplanck\ orientations \citep{kobulnicky1994,falceta2008,hildebrand2009}.

This paper is organized as follows. 
Sect.~\ref{section:data} introduces the \Planck\ $353\,$GHz polarization maps, the \nh\ maps, and the starlight polarization observations. 
Sect.~\ref{section:GaussianRealization} introduces the Gaussian polarization models that we use to evaluate the results the analysis.
Sect.~\ref{section:analysis} describes the vicinity and the \stwo\ statistical analyses performed to compare submillimetre and starlight polarization data.
In Sect.~\ref{section:discussion} we discuss the results of the comparison between both techniques and Sect.~\ref{section:conclusions} summarizes the main results.
Finally, Appendix \ref{section:HRO} adds a commentary on the trends in relative orientations between \bperpstars\ and \nh\ structures.

\begin{table*}%[tmb]  % table* is a two-column table.  Drop the * for one column.
\begingroup
\newdimen\tblskip \tblskip=5pt
\caption{Locations of the selected regions and properties of the starlight polarization observations}
\label{table-fields}                            % Label goes here.
\nointerlineskip
\vskip -3mm
\footnotesize
\setbox\tablebox=\vbox{
   \newdimen\digitwidth 
   \setbox0=\hbox{\rm 0} 
   \digitwidth=\wd0 
   \catcode`*=\active 
   \def*{\kern\digitwidth}
   \newdimen\signwidth 
   \setbox0=\hbox{+} 
   \signwidth=\wd0 
   \catcode`!=\active 
   \def!{\kern\signwidth}
\halign{\hbox to 0.75in{#\leaderfil}\tabskip 2.2em&
\hfil#&
\hfil#&
\hfil#&
\hfil#&
\hfil#&
\hfil#&
\hfil#&
\hfil#\tabskip 0pt\cr
\noalign{\doubleline}
\omit\hfil Region\hfil& \hfil$d$\,$^{a}$\hfil & \hfil$l$\hfil & \hfil$b$\hfil & \hfil$\Delta l \times \Delta b$\hfil & \hfil $N_{\rm star}$$^{b}$ \hfil & Band & \hfil $\min(p_{\rm star}/\sigma_{p_{\rm star}})$\hfil & Reference \cr
\omit&\hfil[pc]\hfil&\hfil[deg]\hfil&\hfil[deg]\hfil&\hfil[deg]\hfil & \hfil \hfil & \hfil \hfil \cr
\noalign{\vskip 4pt\hrule\vskip 6pt}
%----------------------------------------------------------------------------------------------------------------
Taurus & 135 & 172.5 & $-$14.5 & 9.0 $\times$ 9.0 & \phantom{0}287\hfil & $H$ & 3.0\hfil  & \cite{clemens2007}  \cr
            &        &           &              &                            & \phantom{0}474\hfil & $I$  & 5.0\hfil  & \cite{heiles2000}  \cr
%----------------------------------------------------------------------------------------------------------------
Pipe & 145 & 0.0 & 5.0 & 8.0 $\times$ 8.0 & 9796\hfil & $R$ & 5.0\hfil  & \cite{franco2010}  \cr
%----------------------------------------------------------------------------------------------------------------
Lupus I & 140 & 339.0 & 16.0 & 6.0 $\times$ 6.0 & 1938\hfil & $R$ & 5.0\hfil  & \cite{franco2015}  \cr
%----------------------------------------------------------------------------------------------------------------
Musca & 160 & 301.0 & $-$9.0 & 6.0 $\times$ 6.0 & 2439\hfil & $V$ & 5.0\hfil  & \cite{pereyra2004} \cr
%----------------------------------------------------------------------------------------------------------------
\noalign{\vskip 3pt\hrule\vskip 4pt}}}
%\endPlancktable                    % ends one-column \halign
\endPlancktablewide                 % ends two-column \halign
% DISTANCES DISTANCES DISTANCES 
\tablenote a The estimates of distances are from: \cite{schlafly2014} for Taurus, Lupus I, and Musca; \cite{alves2007} for Pipe.\par
\tablenote b Number of stars after thresholding in SNR.\par
%
%\tablenote b Number of selected stars in each region.\par
\endgroup
\end{table*}  

\begin{figure*}[ht!]
\vspace{-0.1cm}
\centerline{
\includegraphics[width=0.49\textwidth,angle=0,origin=c]{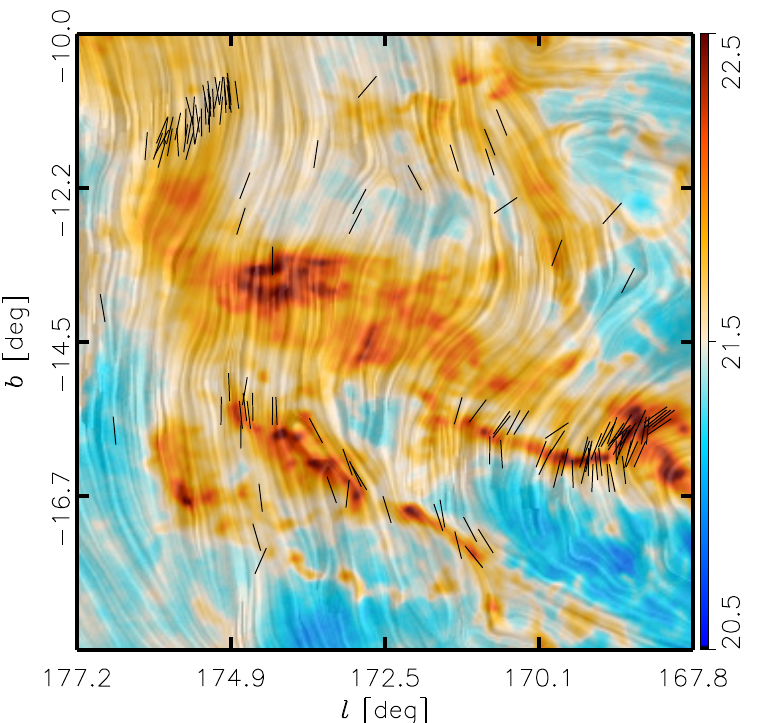}
\includegraphics[width=0.49\textwidth,angle=0,origin=c]{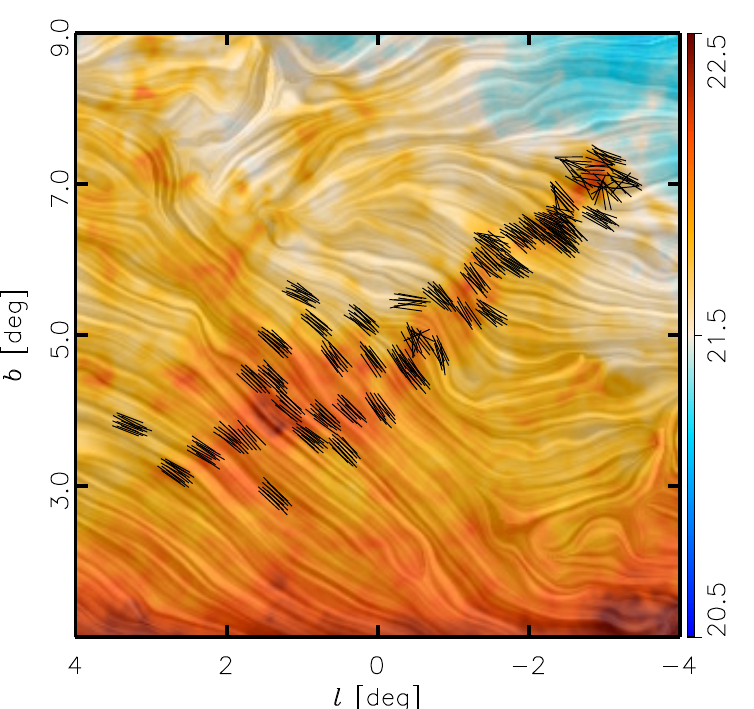}
}
\vspace{-0.2cm}
\centerline{
\includegraphics[width=0.49\textwidth,angle=0,origin=c]{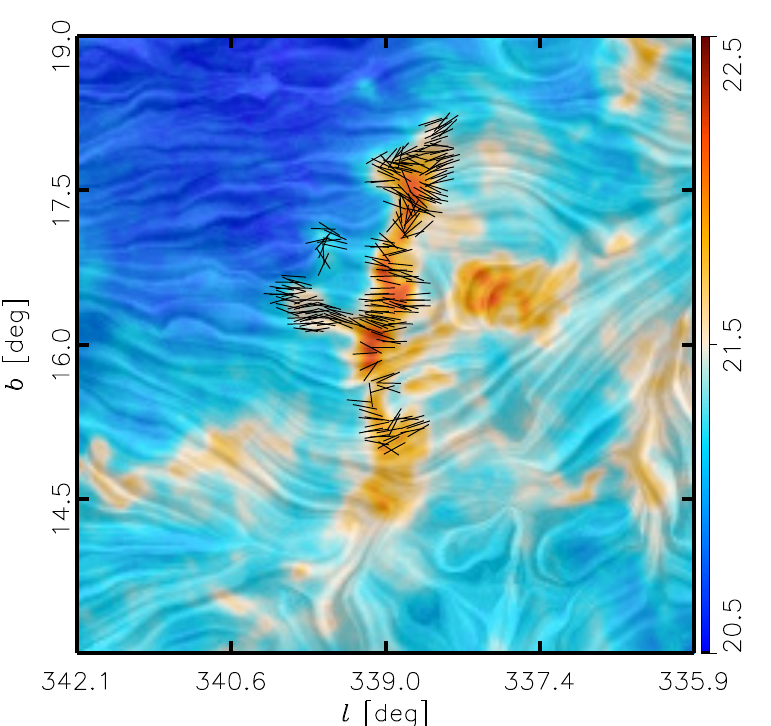}
\includegraphics[width=0.49\textwidth,angle=0,origin=c]{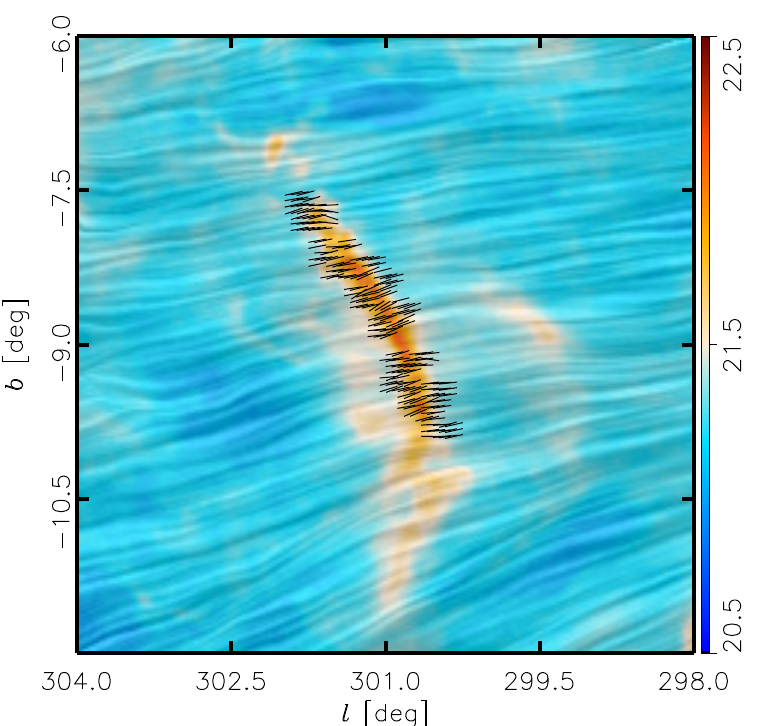}
}
\vspace{-0.2cm}
\caption{Magnetic field orientations inferred from submillimetre emission and visible/NIR extinction polarization observations towards the Taurus {\it (top left)}, Pipe {\it (top right)}, Lupus I {\it (bottom left)}, and Musca {\it (bottom right)} molecular clouds. The colours represent the total gas column density in logarithmic scale. The ``drapery" pattern, produced using the line integral convolution \citep[LIC,][]{cabral1993}, indicates the magnetic field orientation, orthogonal to the orientation of the submillimetre polarization. 
The black pseudo-vectors indicate the magnetic field orientation from starlight polarization in the corresponding lines-of-sight. Each pseudo-vector represents the average field orientation inferred from the stars within 3\parcm5-diameter vicinities.
}
\label{fig:StarlightandLICsubmm}
\end{figure*}

% ==========================================================================================================================
\section{Data}\label{section:data}

\subsection{Thermal dust polarization}\label{data:submmpol}
Over the whole sky, \Planck\ observed linearly polarized emission (Stokes $Q$ and $U$) in seven frequency bands from 30 to 353$\,$GHz \citep{planck2013-p01}. In this study, we used data from the High Frequency Instrument \citep[HFI,][]{lamarre2010} at 353\,GHz, the highest frequency band that is sensitive to polarization.
Towards MCs, the contribution of the cosmic microwave background (CMB) polarized emission is negligible at 353\,GHz, making this \Planck\ map the best suited one to study the spatial structure of dust polarization \citep{planck2014-XIX,planck2014-XX}.

We used the Stokes $Q$ and $U$ maps and the associated noise maps made from five independent consecutive sky surveys of the \Planck\ cryogenic mission, which together correspond to the \Planck\ 2015 public data release\footnote{\url{http://pla.esac.esa.int/pla/}}~\citep{planck2014-a01}. 
The whole-sky 353\,GHz maps of $Q$, $U$, their respective variances $\sigma^{2}_{\textsc{Q}}$, $\sigma^{2}_{\textsc{U}}$, and their covariance $\sigma_{\textsc{QU}}$ are initially at 4\parcm8 resolution in \healpix\ format\footnote{\url{http://healpix.sf.net}}\citep{gorski2005} with a pixelization at $N_{\rm side} = 2048$, which corresponds to an effective pixel size of 1\parcm7. To increase the signal-to-noise ratio (S/N) of extended emission, we smoothed all the maps to 10\arcmin\ resolution using a Gaussian approximation to the \Planck\ beam and the smoothing procedures for the covariance matrix described in \cite{planck2014-XIX}.

The maps of the individual regions are projected and resampled onto a Cartesian grid with the gnomonic projection procedure described in \cite{paradis2012}. The present analysis is performed on these projected maps. The selected regions are small enough, and are located at sufficiently low Galactic latitudes that this projection does not impact significantly on our study.

The Stokes parameters provided by \planck\ data follow the \healpix\ angle convention \citep{gorski2005}, where the polarization angle is measured from the local direction to the north Galactic pole with positive values increasing towards the west (decreasing Galactic longitude $l$). In order to compare these values with starlight polarization observations, we have calculated the polarization angle $\phi$ using the IAU convention \citep{hamaker1996III}, where it is measured from the local direction to the north Galactic pole with positive values increasing towards the east (increasing Galactic longitude $l$). This corresponds to
\begin{equation}\label{qutoangle}
\phi = 0.5 \arctan(-U,Q),
\end{equation}
where the $\arctan(-U, Q)$ function is used to compute $\arctan(-U/Q)$ avoiding the $\pi$ ambiguity.

For this study we assume that the angle of optical/NIR starlight polarization, $\phi_{\rm star}$, is directly equal to the orientation $\psi_{\rm star}$ of \bperp, so that $\psi_{\rm star}=\phi_{\rm star}$. Likewise, we assume that the submillimetre polarization is perpendicular to the orientation of the field, so its polarization angle $\phi_{\rm submm}$ is related to the orientation of the field $\psi_{\rm submm}$ by $\psi_{\rm submm}=\phi_{\rm submm}+\pi/2$.
We use this approximation, implicitly assuming that both polarization observations are homogeneously sampling the magnetic field along the LOS, but it is not necessarily the case. 
The observables in emission and extinction, the Stokes parameters $Q$ and $U$, do not directly trace the magnetic field direction, but rather the density-weighted magnetic field orientation.
Additionally, the alignment of the dust with the local magnetic field is not perfect in all environments \citep{lazarian2007,andersson2015}. 
However, for the sake of comparison between the observed quantities in emission and extinction, the aforementioned approximation is sufficient to compare how both techniques are sampling the ISM.

\subsection{Column density}\label{data:columndensity}
We use the dust optical depth at 353\,GHz ($\tau_{353}$) as a proxy for the total gas column density (\nh). 
The $\tau_{353}$ map was derived from the all-sky \Planck\ intensity observations at 353, 545, and 857$\,$GHz, and the \IRAS\ observations at 100$\,\mu$m, through a modified black body spectrum fit, which also yielded maps of the dust temperature and of the dust opacity spectral index~\citep{planck2013-p06b}.
The $\tau_{353}$ map, computed initially at 5\arcmin\ resolution, was smoothed to 10\arcmin\ to match the polarization maps. The errors resulting from smoothing the product $\tau_{353}$ map, rather than the underlying data and re-fitting, are negligible compared to the uncertainties in the dust opacity and do not significantly affect the results of this study.

To scale from $\tau_{353}$ to \nh, following \cite{planck2013-p06b}, we adopted the dust opacity, 
\begin{equation}\label{eq:nhmap}
\sigma_{353}= \tau_{353}/\nhd = 1.2 \times 10^{-26}\,\mbox{cm}^{2}\,.
\end{equation}
Variations in dust opacity are present even in the diffuse ISM and the opacity decreases systematically by a factor of 2 from the denser to the diffuse ISM \citep{planck2011-7.12,martin2012,planck2013-p06b}, but our results do not depend on this calibration.
  
\subsection{Starlight polarization}\label{data:starlight}
In this analysis, we combine previously published starlight polarization measurements summarized in Table~\ref{table-fields}. We evaluate the observations in terms of their polarized intensity $P\equiv\sqrt{Q^2+U^2}$, only selecting sources in these catalogs with a polarization SNR $P/\sigma_{P} \geq 5$, except for the $H$-band (1.6\micron) polarization observations towards Taurus, where we use all sources with $P/\sigma_{P} \geq 3$. 
These values of polarization SNR correspond to classical uncertainties in the orientation angle $\sigma_{\psi} <$ 5\pdeg7 and $\sigma_{\psi} <$ 9\pdeg5, and they guarantee that the polarization bias is negligible \citep{serkowski1958,naghizadeh-khouei1993,montier2015}.

The optical data towards the Pipe Nebula, Lupus I, and Musca were acquired with the 1.6\,m and 0.6\,m telescopes of the Observat\'orio do Pico dos Dias (LNA/MCTI, Brazil).
The observations were made using IAGPOL, a polarimetric unit consisting of a half-wave plate retarder followed by a calcite Savart prism and a filters wheel \citep{magalhaes1996}. 
The field-of-view of these observations is around 10\arcmin$\times$10\arcmin. 
The data processing provides the reduced Stokes parameters, $q\equiv Q/I$ and $u\equiv U/I$; polarization fraction, $p_{\rm star} \equiv \sqrt{q^2 + u^2}$; and polarization angle, $\phi_{\rm star}$, measured eastwards from the North Celestial Pole (in degrees) for each star.
A detailed description of the polarimetric observations and the data processing can be found in \citep{franco2010} for the Pipe Nebula, \citep{franco2015} for Lupus I, and \citep{pereyra2004} for Musca.

Towards Taurus, the $H$-band polarization was observed with the \emph{Mimir} instrument \citep{clemens2007} using the 1.8\,m Perkins telescope of the Lowell Observatory. 
\emph{Mimir} also contains a rotating half-wave plate and the field of view is the same as the optical observations. 
The $H$-band observations were centred on two subregions: a low-density field and a filamentary, high-density field. 
To cover scales larger than the size of the two subregions but smaller than their separation, we also included in this analysis 474 I-band polarization observations from the catalog described in \cite{heiles2000}, all with polarization SNR $P/\sigma_{P} \geq 5$.

% ============================================================================================================================
\section{Gaussian models of polarization}\label{section:GaussianRealization}

Throughout this work, we characterize our analysis tools using a series of Gaussian models of Stokes $Q$ and $U$ introduced in \cite{planck2014-XXXII} and \cite{planck2016-XLV}.

Each model is built from 3D vectors $\vec{B}_{\rm M} $ with a Gaussian distribution of orientations about a mean direction $\vec{B}_{\rm M0}$. 
The three components of $\vec{B}_{\rm M}$  are independent realizations of a Gaussian field on a full-sky \healpix\ grid, with an angular power spectrum
having a power law, $P(k) \propto k^{\alpha_{\rm M}}$, to which we add the components of $\vec{B}_{\rm M0}$. 
These realizations are computed with the procedure {\tt{SYNFAST}} of \healpix\ at $N_{\rm side}=2048$, which corresponds to an effective pixel size of 1\parcm72. 
%The mean of $\vec{B}_{\rm M}$ is $\vec{B}_{\rm M0}$.

These models are characterized by the ratio $f_{\rm M}$ between the standard deviation of $|\vec{B}_{\rm M}|$ and $|\vec{B}_{\rm M0}|$ and the values of the spectral index, $\alpha_{\rm M}$.
The ratio $f_{\rm M}$ determines the amplitude of the scatter of $\vec{B}_{\rm M}$ with respect to $\vec{B}_{\rm M0}$, while 
the spectral index $\alpha_{\rm M}$ controls the correlation of the $\vec{B}_{\rm M}$ orientations across the sky.
The distribution function of angles between $\vec{B}_{\rm M}$ and $\vec{B}_{\rm M0}$ per solid angle unit is close to Gaussian with a standard deviation, $\sigma_{\rm M}$, that increases from $9\pdeg7$ to $29\pdeg5$ and $38^\circ$ for $f_{\rm M}=0.3$, $1.0$ and $1.5$, respectively.

For each model, we compute maps of the projections of $\vec{B}_{\rm M}$ and $\vec{B}_{\rm M0}$ onto the sky with respect to the local direction of the north Galactic pole, $\psi_{\rm M}$ and $\psi_{\rm M0}$, respectively. 
Subsequently, we compute Stokes $Q_{\rm M} \equiv \sin^{2}\psi_{\rm M} - \cos^{2}\psi_{\rm M}$ and $U_{\rm M}\equiv 2\sin\psi\cos\psi_{\rm M}$, and project them onto a Cartesian-grid map using the procedure described in Section~\ref{data:submmpol}.
These models do not include any density structure and consequently there is no information in the Stokes $I_{\rm M}$.

We chose four models corresponding to the combinations of $\alpha_{\rm M}=-2.5$ and $-1.5$, and $f_{\rm M}=1.0$ and $0.5$. 
We project them towards a $6\deg\times6\deg$ region where the inclination of $\vec{B}_{\rm M0}$ with respect to the plane of the sky is $\gamma=20$\deg.
Given that the mean field inclination in the studied regions is unknown, this selection of $\gamma$ is arbitrary, but it allows to illustrate the effect of the magnetic field structure in the observed polarization.

We analyze each one of these projected models at two angular resolutions, 10\arcmin\ and 2\arcmin, aiming to characterize one of the differences between the starlight and the submillimetre observations.
In reality the angular resolution of each starlight polarization observation is comparable to the size of the star, which is of the order of fractions of an arc-second. 
However, reproducing such a large dynamic range, between the size of the stars and the size of the \Planck\ beam, is unpractical and unnecessary given the fact that we are using a simple model, where most of the structure is in the largest scales.

\begin{figure}[ht!]
\vspace{-0.2cm}
\centerline{
\includegraphics[width=0.45\textwidth,angle=0,origin=c]{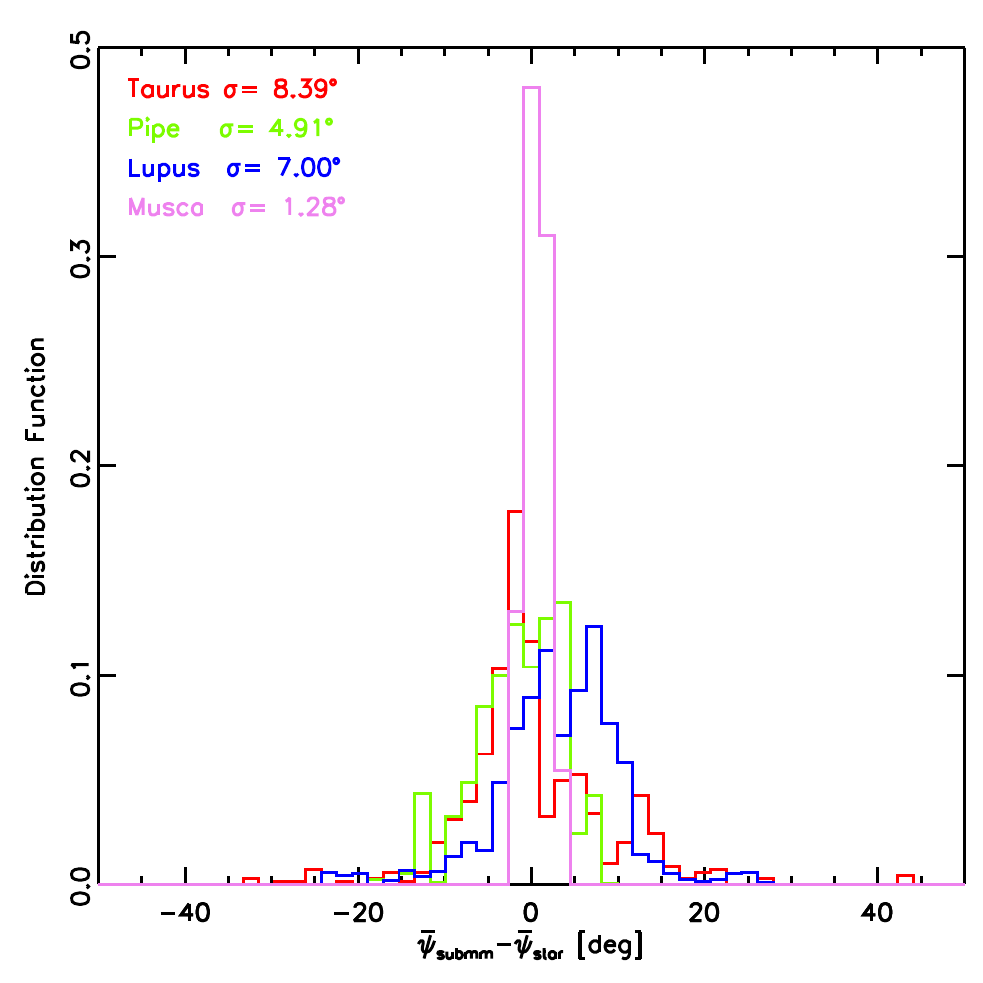}
}
\vspace{-0.2cm}
\caption{Distributions of the differences $\,\bar{\psi}_{\rm submm} - \bar{\psi}_{\rm star}\,$ between the mean orientation of \bperpplanck\ and \bperpstars\ within each 10\arcmin\ vicinity, for each field. The dispersions $\sigma$ of the histograms are given in the top left corner.}
\label{fig:StarAnglevsSubmmAngle}
\end{figure}

\begin{table*}  % table* is a two-column table.  Drop the * for one column.
\begingroup
\newdimen\tblskip \tblskip=5pt
\caption{Field dispersion within 10\arcmin\ vicinities.}
\label{table-means}                            % Label goes here.
\nointerlineskip
\vskip -3mm
\footnotesize
\setbox\tablebox=\vbox{
   \newdimen\digitwidth 
   \setbox0=\hbox{\rm 0} 
   \digitwidth=\wd0 
   \catcode`*=\active 
   \def*{\kern\digitwidth}
   \newdimen\signwidth 
   \setbox0=\hbox{+} 
   \signwidth=\wd0 
   \catcode`!=\active 
   \def!{\kern\signwidth}
\halign{\hbox to 1.15in{#\leaderfil}\tabskip 2.2em&
\hfil#&\hfil#&\hfil#&\hfil#&\hfil#&\hfil#\tabskip 0pt\cr
\noalign{\doubleline}
\omit\hfil Region\hfil & \multispan3\hfil $\bar{\psi}_{\rm star}-\bar{\psi}_{\rm submm}$\hfil & \multispan3\hfil $\varsigma_{\psi_{\rm star}}$ \hfil \cr
\omit \hfil \hfil & \hfil Mean \hfil & \hfil $P_{85}$$^a$ & \hfil $P_{95}$$^a$ \hfil & \hfil Mean \hfil &  \hfil Median  \hfil &  \hfil Std. Dev. \hfil \cr
%\omit\hfil Region\hfil & \hfil $<\varsigma_{\psi_{\rm star}}>$ \hfil &  \hfil $\tilde{\varsigma}_{\psi_{\rm star}}$$^a$  \hfil & \hfil$<\bar{\psi}_{\rm star}-\bar{\psi}_{\rm submm}>$  \hfil & \hfil  max$(\bar{\psi}_{\rm star}-\bar{\psi}_{\rm submm})$  \hfil & \hfil $S^{\rm star}_{2}(10\arcmin)$ \hfil & \hfil $S^{\rm star}_{2}(10\arcmin)-S^{\rm submm}_{2}(10\arcmin)$ \hfil \cr
\omit & \hfil[deg]\hfil & \hfil[deg]\hfil & \hfil[deg]\hfil & \hfil[deg]\hfil & \hfil[deg]\hfil & \hfil[deg]\hfil \cr
\noalign{\vskip 4pt\hrule\vskip 6pt}
%----------------------------------------------------------------------------------------------------------------
%Taurus & \hfil10.4\hfil & \hfil\phantom{0}9.5\hfil & \hfil-0.2\hfil & \hfil28.5\hfil & \hfil18.2\hfil & \hfil\phantom{0}9.1\hfil \cr
Taurus & \hfil-0.2\hfil & \hfil12.0\hfil & \hfil16.6\hfil & \hfil10.4\hfil & \hfil\phantom{0}9.5\hfil & \hfil6.1\hfil \cr
%----------------------------------------------------------------------------------------------------------------
%Pipe & \hfil\phantom{0}7.5\hfil & \hfil\phantom{0}6.0\hfil & \hfil\phantom{0}1.1\hfil & \hfil\phantom{0}9.1\hfil & \hfil11.1\hfil & \hfil\phantom{0}3.9\hfil\cr
Pipe & \hfil\phantom{0}1.1\hfil & \hfil\phantom{0}6.9\hfil & \hfil12.4\hfil & \hfil\phantom{0}7.5\hfil & \hfil\phantom{0}6.0\hfil & \hfil5.3\hfil \cr
%----------------------------------------------------------------------------------------------------------------
Lupus I & \hfil\phantom{0}4.4\hfil & \hfil10.8\hfil& \hfil15.2\hfil & \hfil19.0\hfil & \hfil15.2\hfil & \hfil9.9\hfil \cr
%----------------------------------------------------------------------------------------------------------------
%Musca & \hfil\phantom{0}6.2\hfil & \hfil\phantom{0}6.1\hfil & \hfil\phantom{0}1.5\hfil & \hfil\phantom{0}5.1\hfil & \hfil\phantom{0}9.8\hfil & \hfil\phantom{0}5.9\hfil\cr
Musca & \hfil\phantom{0}1.5\hfil & \hfil\phantom{0}2.7\hfil & \hfil\phantom{0}3.7\hfil & \hfil\phantom{0}6.2 & \hfil\phantom{0}6.1\hfil & \hfil1.0\hfil  \cr
%----------------------------------------------------------------------------------------------------------------
\noalign{\vskip 3pt\hrule\vskip 4pt}}}
\endPlancktablewide                    % ends one-column \halign
\tablenote a Percentile values of $|\,\bar{\psi}_{\rm submm} - \bar{\psi}_{\rm star}\,|$.\par
\endgroup
\end{table*}  
\begin{figure*}[ht!]
\vspace{-0.2cm}
\centerline{
\includegraphics[width=0.45\textwidth,angle=0,origin=c]{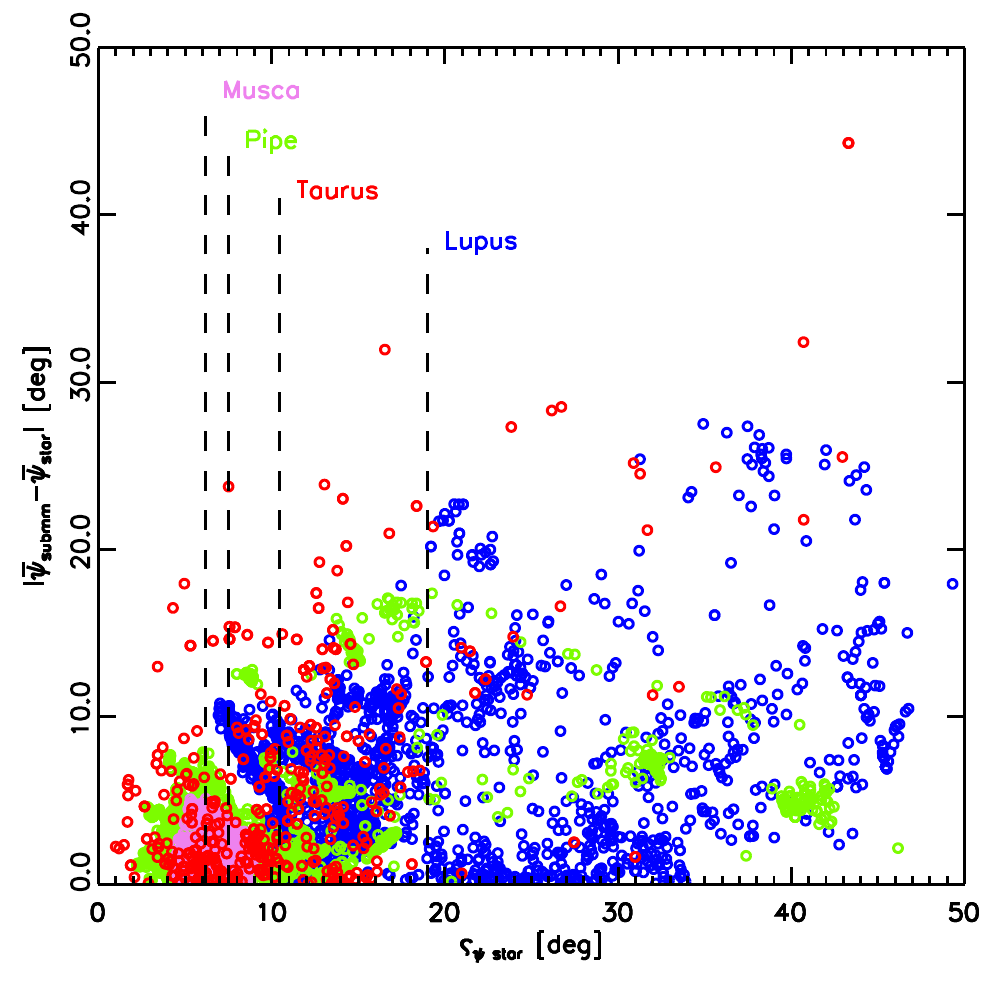}
\includegraphics[width=0.45\textwidth,angle=0,origin=c]{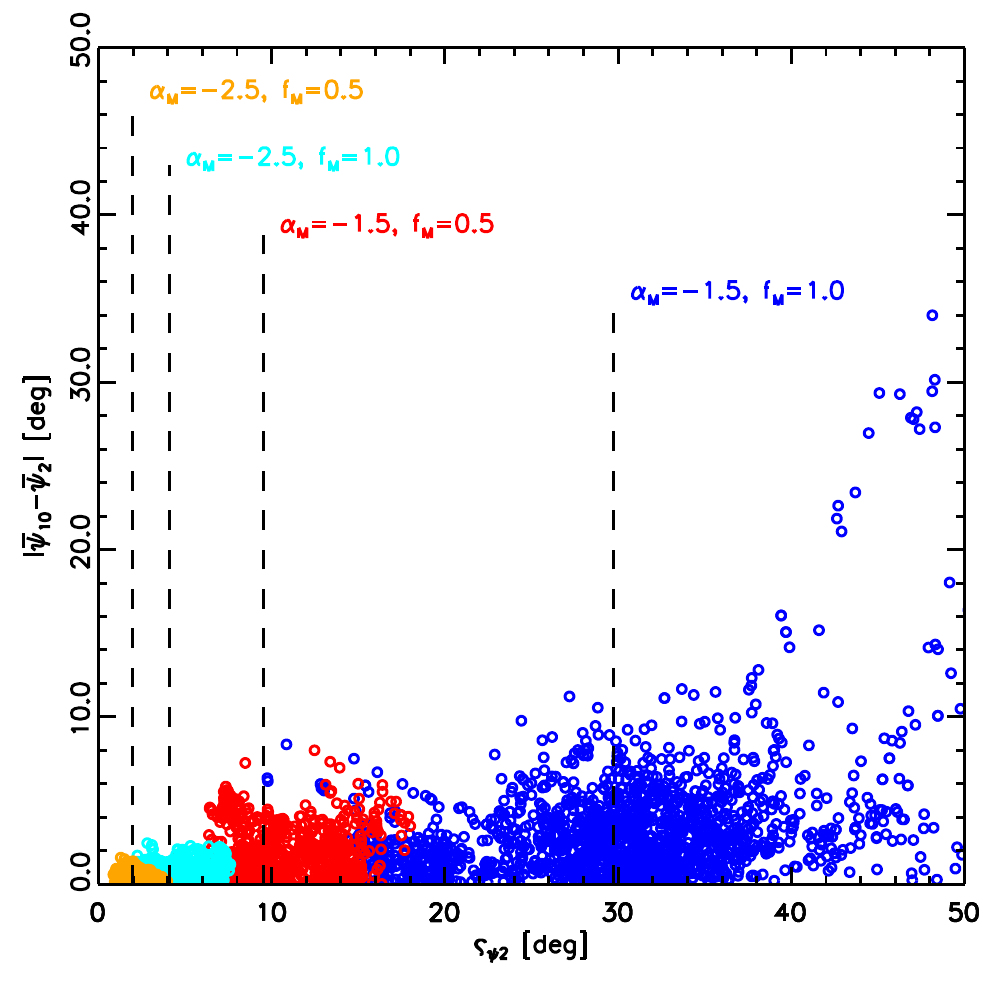}
}
\vspace{-0.3cm}
\caption{Scatter plot of the differences between the mean field orientations inferred from starlight and submillimeter polarization, $|\,\bar{\psi}_{\rm submm} - \bar{\psi}_{\rm star}\,|$, against the 
dispersion of orientation angles, $\varsigma_{\psi_{\rm star}}$, in all the 10\arcmin\ vicinities with more than three stars towards each observed region (\emph{left}) and towards the Gaussian polarization models introduced in Sect.~\ref{section:GaussianRealization} (\emph{right}). The dashed vertical lines indicate the mean values of $\varsigma_{\psi_{\rm star}}$.}
\label{fig:DiffVsSigmaPsi}
\end{figure*}
\begin{figure*}[ht!]
\vspace{-0.2cm}
\centerline{
\includegraphics[width=0.45\textwidth,angle=0,origin=c]{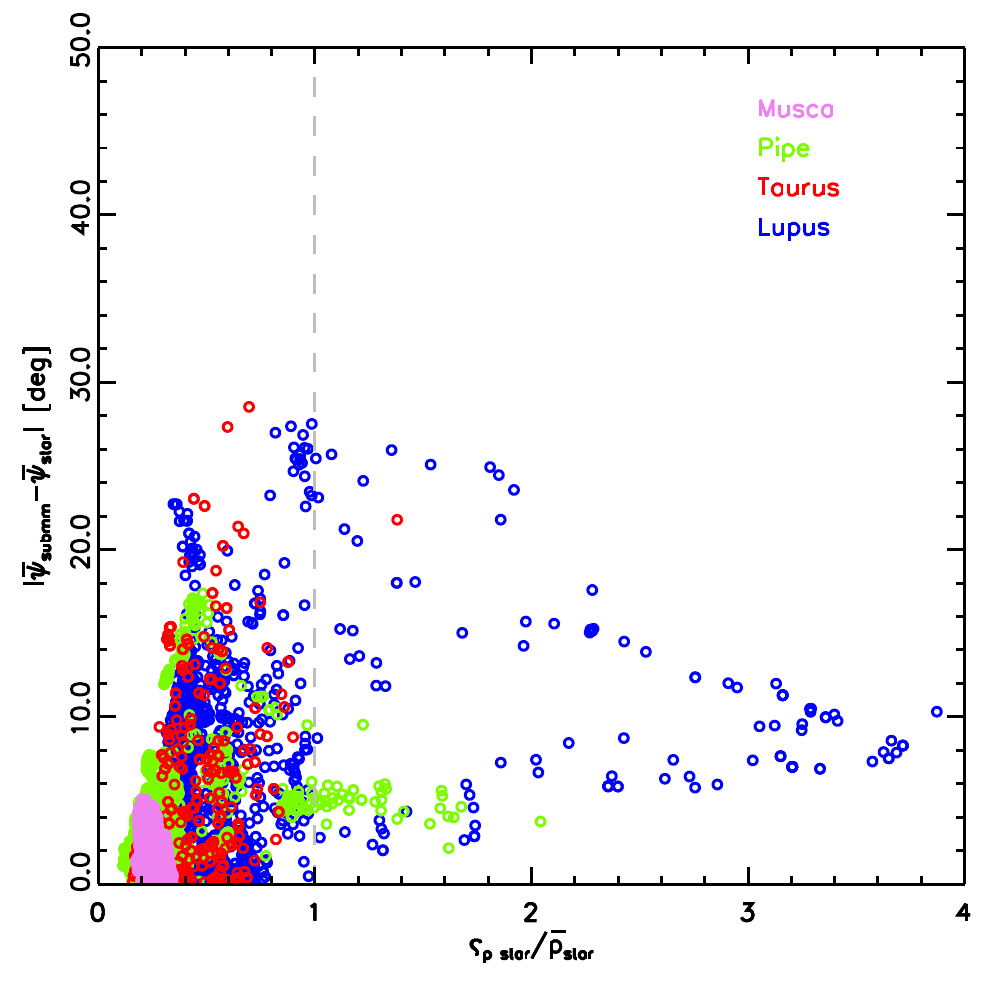}
\includegraphics[width=0.45\textwidth,angle=0,origin=c]{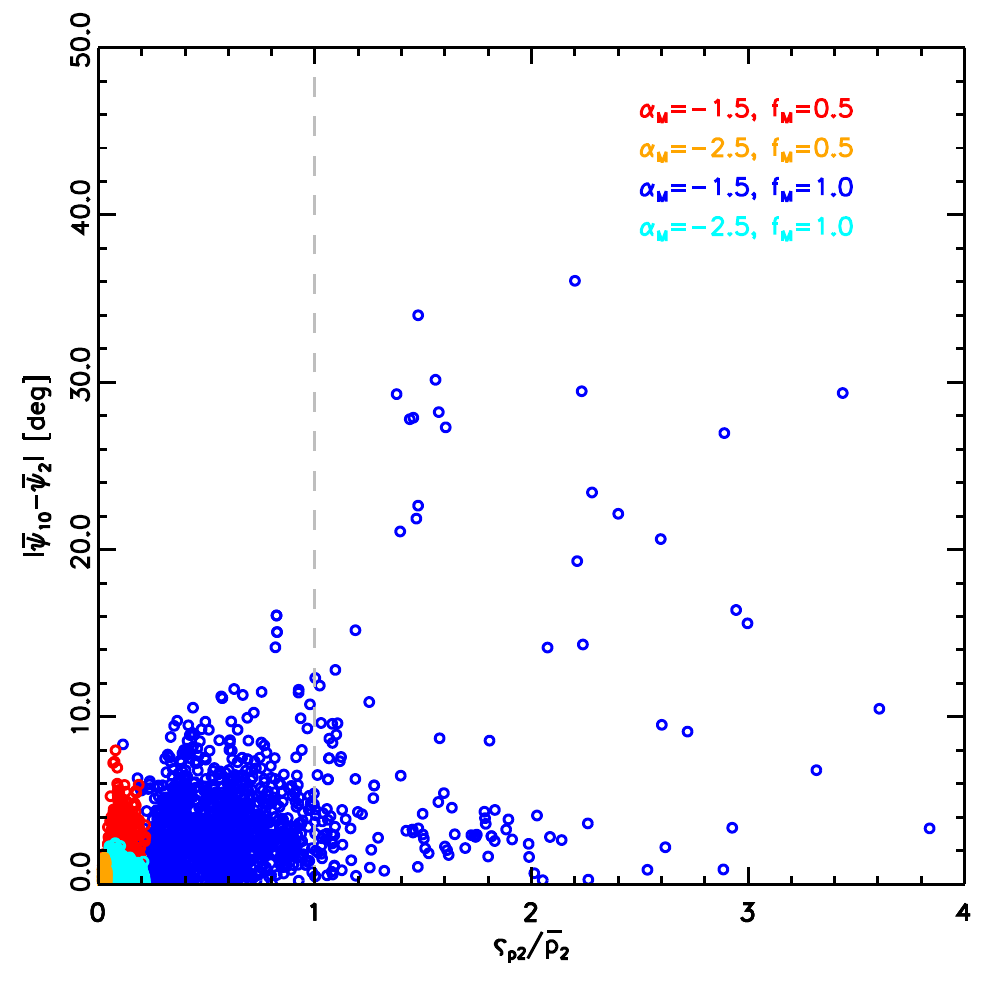}
}
\vspace{-0.3cm}
\caption{Scatter plot of the differences between the mean field orientations inferred from starlight and submillimeter polarization, $|\,\bar{\psi}_{\rm submm} - \bar{\psi}_{\rm star}\,|$, against the fluctuations in polarized fraction, $\varsigma_{p_{\rm star}}/\bar{p}_{\rm star}$, in all the 10\arcmin\ vicinities with more than three stars towards each observed region (\emph{left}) and towards the Gaussian polarization models introduced in Sect.~\ref{section:GaussianRealization} (\emph{right}). The dashed grey line indicates $\varsigma_{p_{\rm star}}/\bar{p}_{\rm star}=1$.}
\label{fig:DiffVsSigmaPoverP}
\end{figure*}
\begin{figure*}[ht!]
\vspace{-0.2cm}
\centerline{
\includegraphics[width=0.45\textwidth,angle=0,origin=c]{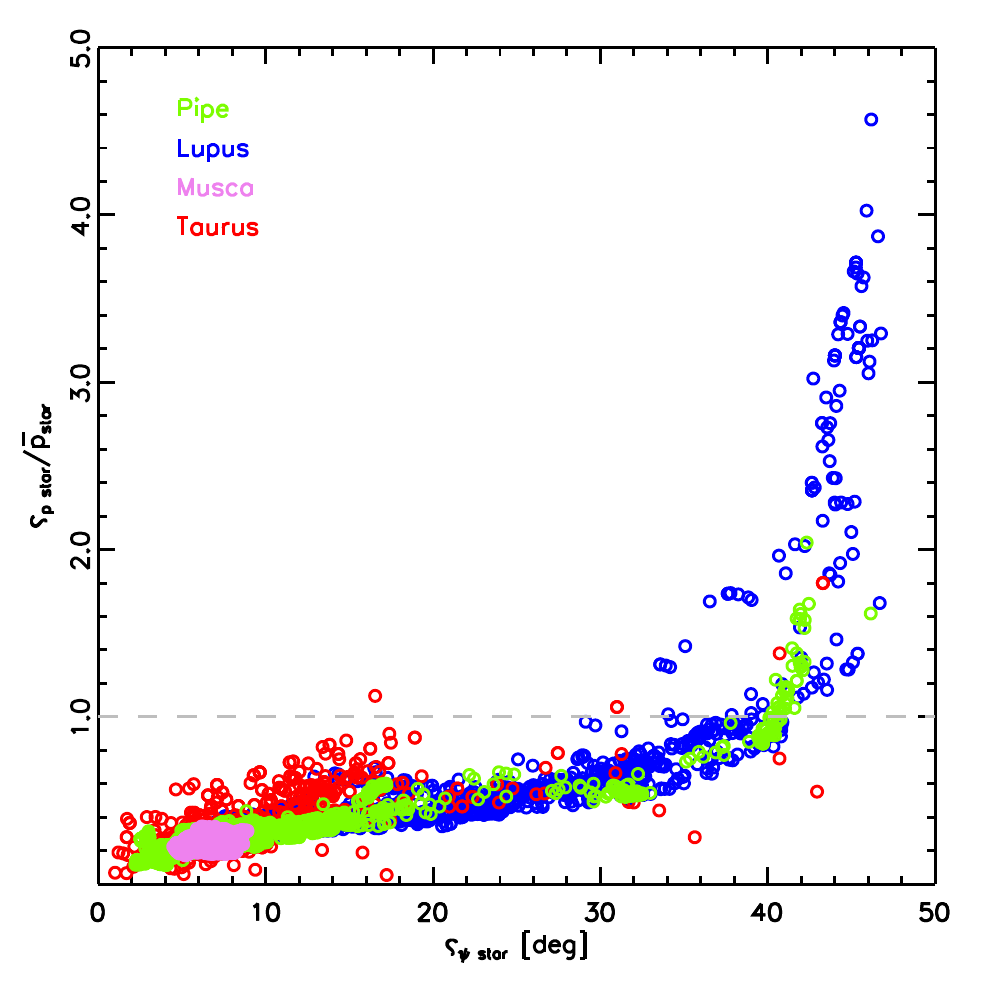}
\includegraphics[width=0.45\textwidth,angle=0,origin=c]{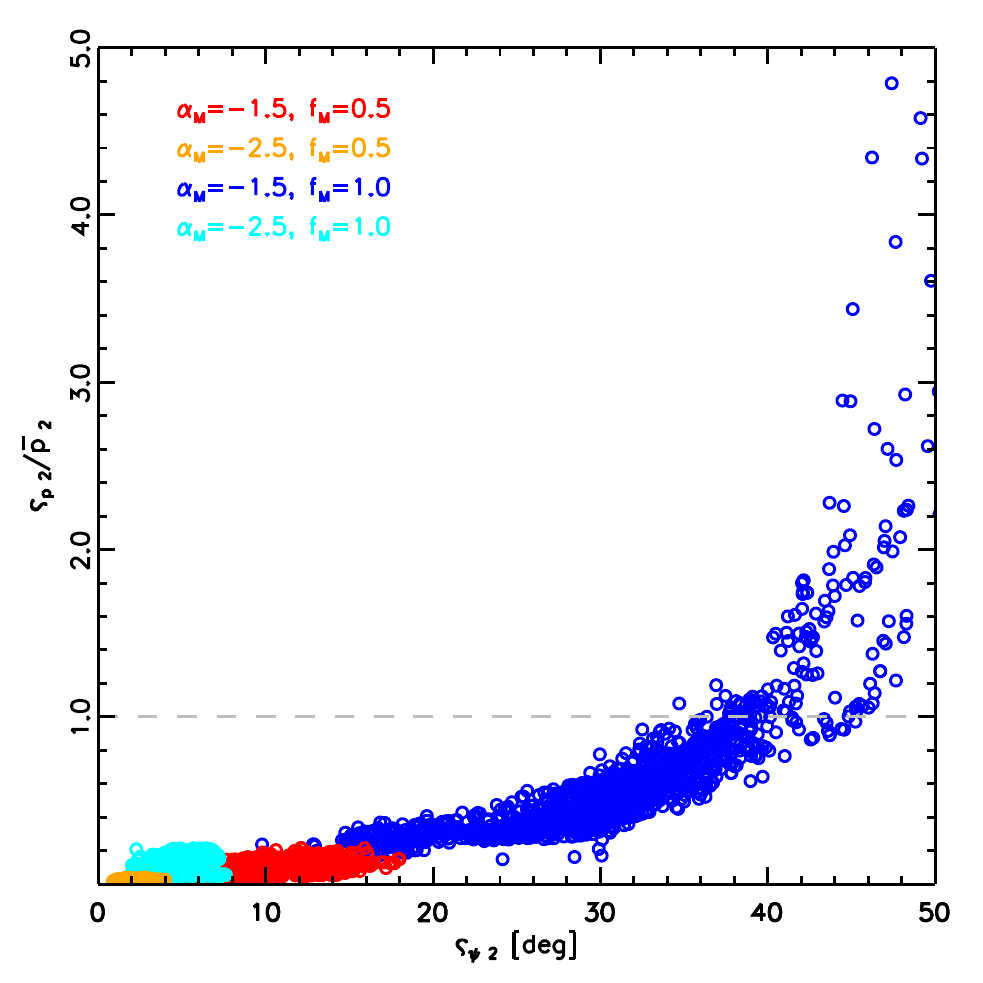}
}
\vspace{-0.3cm}
\caption{Scatter plot of the fluctuations in polarized fraction, $\varsigma_{p_{\rm star}}/\bar{p}_{\rm star}$ against the dispersion of orientation angles, $\varsigma_{\psi_{\rm star}}$, in all the 10\arcmin\ vicinities with more than three stars towards each observed region (\emph{left}) and towards the Gaussian polarization models introduced in Sect.~\ref{section:GaussianRealization} (\emph{right}). The dashed grey line indicates $\varsigma_{p_{\rm star}}/\bar{p}_{\rm star}=1$.}
\label{fig:SigmapOverMeanpVsSigmaPsi}
\end{figure*}

% ==================================================================================================================================
\section{Analysis}\label{section:analysis}
The spatial distribution of $\psi_{\rm star}$ and $\psi_{\rm submm}$ towards the four selected regions is presented in Fig.~\ref{fig:StarlightandLICsubmm}. 
The former is shown as a set of uniform length pseudo-vectors representing the average field orientation inferred from the stars within 3\parcm5-diameter vicinities. 
The latter is shown as a ``drapery'' pattern, produced from the raw \Planck\ 353\,GHz observations using the line integral convolution \citep[LIC,][]{cabral1993}, overlaid on the corresponding \nh\ map. 

In order to quantitatively compare both types of observations, we compute the mean and the dispersion of starlight polarization orientations and polarized fractions within an area identical to that of the submillimetre beam. 
Then, we compute the structure function of the magnetic field orientations to characterize and compare the field structures sampled by each type of observations across multiple scales in each region. 
We present this analysis applied to both the observations and the Gaussian polarization models introduced in Sect.~\ref{section:GaussianRealization}.

% ---------------------------------------------------------------
\subsection{Structure of \bperp\ within the \Planck\ beam}

Towards each studied region, we identify a series of vicinities centred on each of the selected stars and with the same diameter, $d=10$\arcmin, as the \planck\ beam.
For each vicinity containing more than three stars, we evaluate the dispersion of the starlight-inferred magnetic field orientations, $\varsigma_{\psi_{\rm star}} $, 
the difference between the mean field orientations inferred from starlight polarization and \Planck\ 353\,GHz observations, $|\,\bar{\psi}_{\rm star}-\bar{\psi}_{\rm submm}\,|$, 
and the ratio between the dispersion and the mean value of the starlight polarization fraction, $\varsigma_{p_{\rm star}}/p_{\rm star}$.

\subsubsection{Magnetic field orientation within the \Planck\ beam}

We compute the mean magnetic field orientation in a given vicinity
\begin{equation}\label{eq:starmean}
\bar{\psi}_{\rm star} =  \frac{1}{N}\sum\limits_{i=1}^{N} \psi_{\rm star}^{(i)},
\end{equation}
and its dispersion
\begin{equation}\label{eq:stardispersion}
\varsigma_{\psi_{\rm star}} =  \left(\frac{1}{N-1}\sum\limits_{i=1}^{N} (\bar{\psi}_{\rm star} - \psi_{\rm star}^{(i)})^{2} \right)^{1/2}\, ,
\end{equation}
where $N$ is the number of stars in that vicinity. 
We use the notation $\varsigma_{\psi_{\rm star}}$ to avoid confusion with $\sigma_{\psi_{\rm star}}$, 
which is the uncertainty on the orientation angle ${\psi_{\rm star}}$.
The quantity $\bar{\psi}_{\rm star} - \psi_{\rm star}^{(i)}$ is evaluated considering the periodicity of the angles by using
\begin{equation}\label{eq:DeltaPsifromQUstar}
\bar{\psi}_{\rm star} - \psi_{\rm star}^{(i)} = \frac{1}{2}\arctan\left(q^{(i)}_{\rm star}\,\bar{u}_{\rm star}-\bar{q}_{\rm star}\,u^{(i)}_{\rm star}\,, \,q^{(i)}_{\rm star}\,\bar{q}_{\rm star}+u^{(i)}_{\rm star}\,\bar{u}_{\rm star}\right)\, ,
\end{equation}
where $q^{(i)}_{\rm star}$ and $u^{(i)}_{\rm star}$ are the values of the reduced Stokes parameters corresponding to the polarization observation of the $i$-th star, while $\bar{q}_{\rm star}$ and $\bar{u}_{\rm star}$ are the averages of these quantities computed over all the stars in the vicinity,
which correspond to $\bar{\psi}_{\rm star}$ when the polarization bias is small.
 
We computed the difference between the mean orientation of the field derived from starlight and submillimetre by using 
\begin{align}\label{eq:DeltaPsifromQU}
&\bar{\psi}_{\rm star}-\bar{\psi}_{\rm submm} = \\ 
&\frac{1}{2}\arctan\left(\bar{Q}_{\rm submm}\bar{u}_{\rm star}-\bar{q}_{\rm star}\bar{U}_{\rm submm}, \,\bar{Q}_{\rm submm}\bar{q}_{\rm star}+\bar{U}_{\rm submm}\bar{u}_{\rm star}\right)\, , \nonumber
\end{align}
where $\bar{Q}_{\rm submm}$ and $\bar{U}_{\rm submm}$ are the mean values of the \Planck\ 353\,GHz Stokes parameters within the vicinity.
We show the distributions of $\bar{\psi}_{\rm star}-\bar{\psi}_{\rm submm}$ towards each region in Fig.~\ref{fig:StarAnglevsSubmmAngle}.

We compare the estimated values of $|\,\bar{\psi}_{\rm star}-\bar{\psi}_{\rm submm}\,|$ with the dispersion of starlight-inferred magnetic field orientations, $\varsigma_{\psi_{\rm star}}$, within each 10\arcmin\ vicinities in Fig.~\ref{fig:DiffVsSigmaPsi}. 
We present some of the parameters that describe the distributions of $\varsigma_{\psi_{\rm star}}$ and $|\,\bar{\psi}_{\rm star}-\bar{\psi}_{\rm submm}\,|$ in Table~\ref{table-means}.

\subsubsection{Starlight polarized fraction $p_{\rm star}$ within the \Planck\ beam}

We compute the polarization fraction of the mean starlight polarization signal
\begin{equation}\label{eq:starpmean}
\bar{p}_{\rm star} =  \left(\bar{q}^{2}_{\rm star} + \bar{u}^{2}_{\rm star} \right)^{1/2} ,
\end{equation}
and the dispersion
\begin{equation}\label{eq:starpdispersion}
\varsigma_{p_{\rm star}} =  \left(\frac{1}{N-1}\sum\limits_{i=1}^{N}\left[(\bar{q}_{\rm star} - q_{\rm star})^{2} + (\bar{u}_{\rm star} - u_{\rm star})^{2} \right]   \right)^{1/2},
\end{equation}
where $N$ is the number of stars in that vicinity. 
As in the case of $\varsigma_{\psi_{\rm star}}$, we use the notation $\varsigma_{p_{\rm star}}$ to avoid confusion with $\sigma_{p_{\rm star}}$, which is the uncertainty on the polarization fraction ${p_{\rm star}}$.
We compare the estimated values of the $|\,\bar{\psi}_{\rm star}-\bar{\psi}_{\rm submm}\,|$ with the $p_{\rm star}$ fluctuations characterized by $\varsigma_{p_{\rm star}}/\bar{p}_{\rm star}$ in Fig.~\ref{fig:DiffVsSigmaPoverP}.

%Both $p_{\rm star}$ and $\psi_{\rm star}$ come from the same Stokes parameters $q_{\rm star}$ and $u_{\rm star}$, thus they are correlated.
%However, if there is structure inside of the vicinity, the angle dispersion $\varsigma_{\psi_{\rm star}}$ saturates at the maximum value of angle dispersion, $\pi/\sqrt{12}$\,radians $\approx 52$\degr, while $\varsigma_{p_{\rm star}}$ has no upper limit.

% ----------------------------------------------------------------------------------------------------------------------------------------------------------------------------------------------------------------------
% STRUCTURE FUNCTION STRUCTURE FUNCTION STRUCTURE FUNCTION STRUCTURE FUNCTION STRUCTURE FUNCTION STRUCTURE FUNCTION  
% ----------------------------------------------------------------------------------------------------------------------------------------------------------------------------------------------------------------------
\subsection{Angular structure function}\label{subsection:SF}

In order to further characterize the values of $\psi_{\rm star}$ and $\psi_{\rm submm}$ without averaging over a particular vicinity size, we evaluate their second-order structure function, \stwo. 
In this technique, the \bperp\ orientation dispersion is quantified by considering the difference in angle,  $\Delta\psi(\boldsymbol{\ell}) = \psi(\vec{x}) - \psi(\vec{x+\boldsymbol{\ell}})$, between pairs of \bperp\ pseudo-vectors separated by displacements $\boldsymbol{\ell}$ in the plane of the sky. 
Assuming that the angle differences are statistically isotropic (i.e., they depend only on $\ell=|\boldsymbol{\ell}|$ and not on the orientation of $\boldsymbol{\ell}$), they can be binned by distance, $\ell$. 
Considering a central position $\vec{x}$, there are $N(\ell)$ pixels, identified by an index $\vec{i}$, such that the distance from $\vec{x}$ to $\vec{x}_{i}$ lies within the corresponding $\ell$-bin. 
The square of the second-order structure function is then,
\begin{equation}\label{eq:strucfunc}
S^{2}_{2}(\ell) =  \langle\frac{1}{N(\ell)}\sum\limits_{i=1}^{N(\ell)} (\Delta\psi_{x,i})^{2} \rangle_{x}\, ,
\end{equation}
where $\Delta\psi_{x,i}=\psi(\vec{x})-\psi(\vec{x}_i)$ and $\langle\dots\rangle_{x}$ denotes an average over all the selected observations.
In terms of the Stokes parameters, this difference can be written
\begin{equation}\label{eq:dispfunc}
\Delta\psi_{x,i} = \frac{1}{2}\arctan\left(Q_{i}U_x-Q_xU_{i}\,, \,Q_{i}Q_x+U_{i}U_x\right)\, ,
\end{equation}
with $Q_{i}=Q(\vec{x}_{i})$, $U_{i}=U(\vec{x}_i)$, $Q_x=Q(\vec{x})$, and $U_x=U(\vec{x})$.

We calculate \stwo\ for both starlight and submillimetre polarization observations using the lines of sight where starlight observations are available and calculating the corresponding ${\psi}_{\rm submm}$ at the exact position of the stars by linear interpolation of the \Planck\ 353\,GHz Stokes $Q$ and $U$ maps. 
For the sake of comparison, we also compute \stwo\ for 10,000 lines of sight randomly distributed over the \Planck\ 353\,GHz Stokes $Q$ and $U$ maps.

To construct \stwo\, we first compute the angle difference, $\Delta\psi_{x,i}$, for every pair of points in each region.   
Next, we bin the data into 40 bins of equal length between 0 and 120\arcmin.
The values of the lag, $\ell$, correspond to the geometrical average of the distances used to define each bin.
The variance of \stwo\ is calculated by random sampling (bootstrapping).
The calculated values of the angular structure function for the starlight and the submillimetre polarization observations, \stwostar\ and \stwosubmm, respectively, are presented in Fig.~\ref{fig:StructureFunctionZoom}.

\subsection{Differences between \stwostar\ and \stwosubmm}

\begin{figure*}[ht!]
\centerline{
\includegraphics[width=0.48\textwidth,angle=0,origin=c]{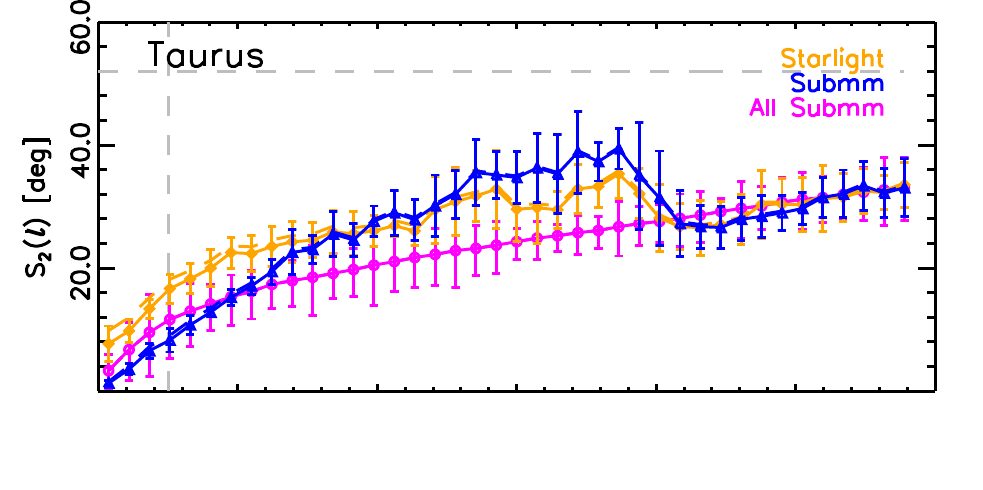}
\includegraphics[width=0.48\textwidth,angle=0,origin=c]{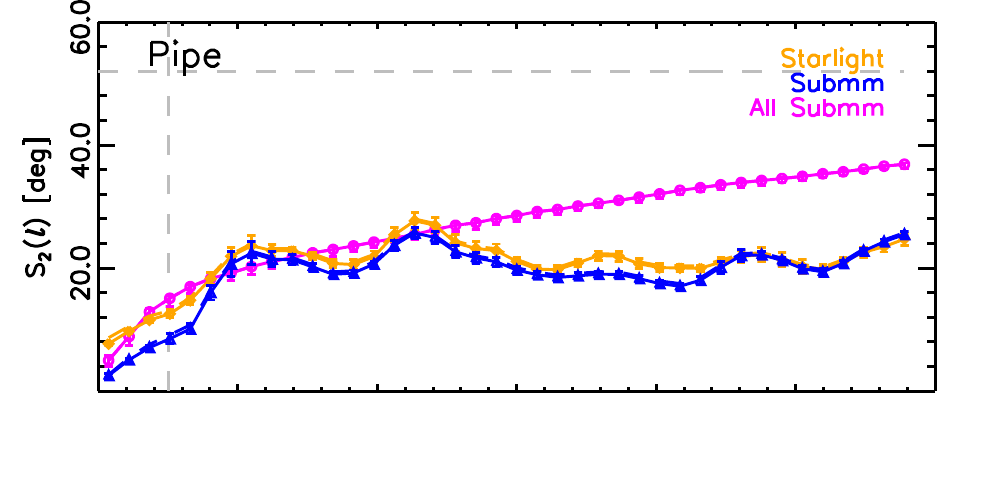}
}
\vspace{-1.0cm}
\centerline{
\includegraphics[width=0.48\textwidth,angle=0,origin=c]{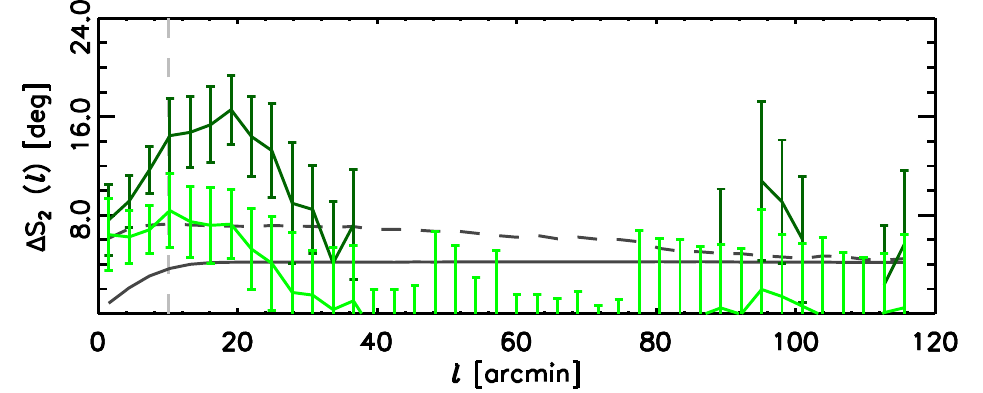}
\includegraphics[width=0.48\textwidth,angle=0,origin=c]{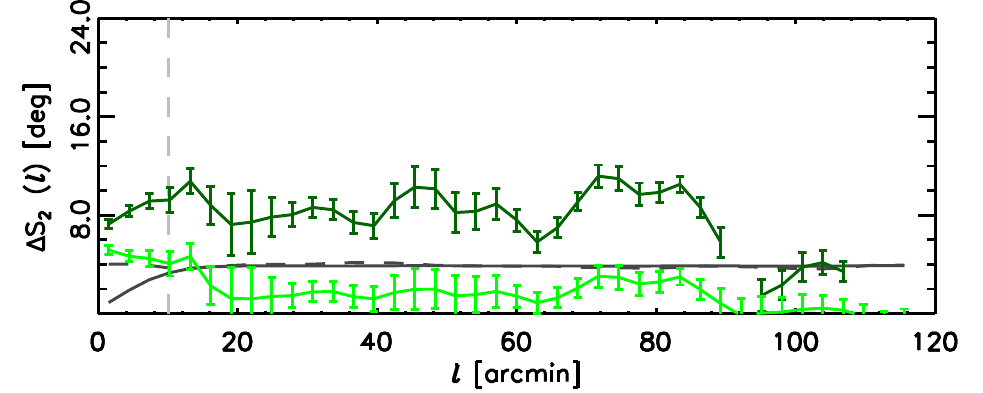}
}
\vspace{-0.1cm} % --------------------------------------------------------------------------------------------------------------------------------------------
\centerline{
\includegraphics[width=0.48\textwidth,angle=0,origin=c]{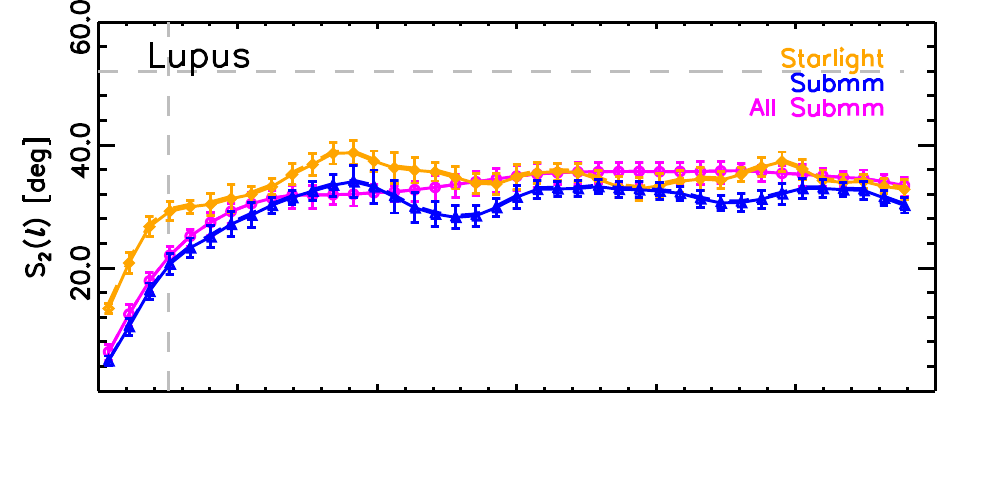}
\includegraphics[width=0.48\textwidth,angle=0,origin=c]{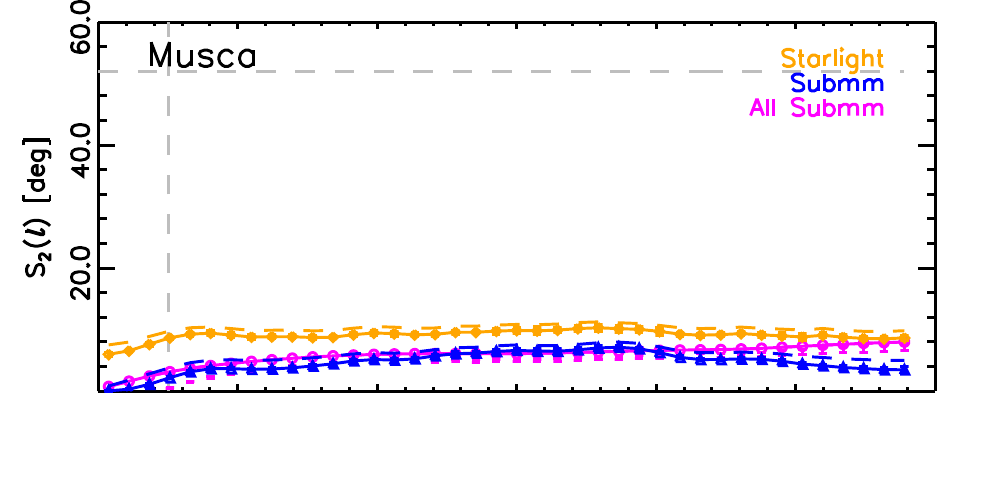}
}
\vspace{-1.0cm}
\centerline{
\includegraphics[width=0.48\textwidth,angle=0,origin=c]{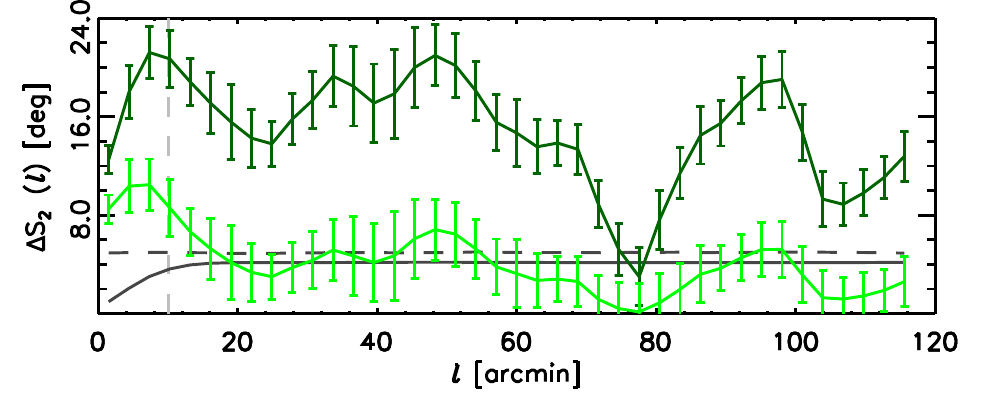}
\includegraphics[width=0.48\textwidth,angle=0,origin=c]{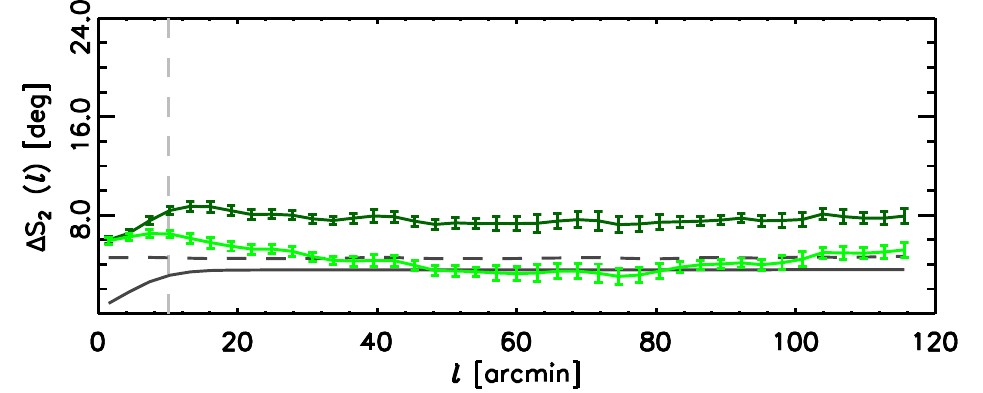}
}
\caption{Structure functions of the starlight polarization, \stwostar\ (orange), submillimetre polarization along the lines of sight to stars, \stwosubmm\ (blue), and submillimetre polarization over the whole map, \stwoallsubmm\ (magenta).
In the lower plot of each panel we present the differences
$\Delta S_{2}(\ell) \equiv \left[([S^{\rm star}_{2}(\ell)]^{2}-C^{2}_{\rm star})-([S^{\rm submm}_{2}(\ell)]^{2}-C^{2}_{\rm submm})\right]^{1/2}$
 (dark green), $\delta S_{2}(\ell) \equiv S^{\rm star}_{2}(\ell)-S^{\rm submm}_{2}(\ell)$ (light green), and the corresponding bias corrections, $C^{2}_{\rm star}$ (dashed grey) and $C^{2}_{\rm submm}$ (solid grey).
The vertical dashed line indicates the effective size of the \planck\ beam. The horizontal dashed line in the upper plot of each panel indicates $S_{2}(\ell)=\pi/\sqrt{12}$\,rad $\approx 52$\degr, which corresponds to a random distribution of orientations. 
}
\label{fig:StructureFunctionZoom}
\end{figure*}

We quantify the differences between \stwostar\ and \stwosubmm\ using two methods. 
First, 
\begin{equation}\label{eq:deltaS2lin}
\delta S_{2}(\ell) \equiv S^{\rm star}_{2}(\ell)-S^{\rm submm}_{2}(\ell), 
\end{equation}
which corresponds to what can be inferred from the visual inspection of the values illustrated in the upper plot of each panel in Fig.~\ref{fig:StructureFunctionZoom}.
Second, 
\begin{equation}\label{eqn:deltaS2}
%\Delta S_{2}(\ell) \equiv \left[([S^{\rm star}_{2}(\ell)]^{2}-C^{2}_{\rm star})-([S^{\rm submm}_{2}(\ell)]^{2}-C^{2}_{\rm submm})\right]^{1/2}, 
\Delta S_{2}(\ell) \equiv \sqrt{([S^{\rm star}_{2}(\ell)]^{2}-C^{2}_{\rm star})-([S^{\rm submm}_{2}(\ell)]^{2}-C^{2}_{\rm submm})}
\end{equation}
which corresponds to the quadratic differences between both functions and accounts for the effects of the bias corrections, $C^{2}(\ell)$.

Although we selected polarization observations with high SNR, we nevertheless evaluate the effect of the polarization bias on the angular structure function, as follows.
Each orientation angle $\psi$ is given by
\begin{equation}
\psi(\vec{x}) = \psi_{0}(\vec{x}) + \delta_\psi(\vec{x}),
\end{equation}
where $ \psi_{0}(\vec{x})$ is the true value of the angle and $\delta_\psi(\vec{x})$ a random (zero-mean) error, then the expectation value of the square of 
\begin{align}
\Delta\psi(\vec{\ell}) &=\psi(\vec{x}) - \psi(\vec{x}+\vec{\ell}) \nonumber \\ 
			       &= [\psi_{0}(\vec{x}) + \delta_\psi(\vec{x})] - [\psi_{0}(\vec{x}+\vec{\ell}) + \delta_\psi(\vec{x}+\vec{\ell})]  \nonumber \\ 
			       &= \Delta\psi_{0}(\vec{\ell}) + [\delta_\psi(\vec{x}) - \delta_\psi(\vec{x}+\vec{\ell})]
\end{align}
is given by
\begin{align}
\langle[\Delta\psi(\vec{\ell})]^2\rangle &= \langle\Delta\psi_0(\vec{\ell})]^2\rangle \nonumber + \langle[\delta_\psi(\vec{x})]^2\rangle \\ 
	& + \langle[\delta_\psi(\vec{x}+\vec{\ell})]^2\rangle - 2\langle\delta_\psi(\vec{x})\delta_\psi(\vec{x}+\vec{\ell})\rangle .
\end{align}
For the starlight polarization observations $\langle\delta_\psi(\vec{x})\delta_\psi(\vec{x}+\vec{\ell})\rangle = 0$, because they correspond to a pencil-like beam and the measurements are uncorrelated, and by definition the mean of the square of the measurement uncertainties, $\sigma^2_\psi(\vec{x})\equiv\langle[\delta_\psi(\vec{x})]^2\rangle$ and $\sigma^2_\psi(\vec{x}+\vec{\ell})\equiv\langle[\delta_\psi(\vec{x}+\vec{\ell})]^2\rangle$, thus leading to
\begin{align}
\langle[\Delta\psi(\vec{\ell})]^2\rangle &= \langle\Delta\psi_0(\vec{\ell})]^2\rangle + \sigma^2_\psi(\vec{x}) + \sigma^2_\psi(\vec{x}+\vec{\ell}).
\end{align}

It follows that in order to recover the true value
\begin{align}
\langle\Delta\psi_0(\vec{\ell})]^2\rangle &= \langle[\Delta\psi(\vec{\ell})]^2\rangle - C^{2}_{\rm star},
\end{align}
we have to subtract the de-biasing correction given by 
\begin{align}\label{eq:CorrStars}
C^{2}_{\rm star} &= \sigma^2_\psi(\vec{x}) + \sigma^2_\psi(\vec{x}+\vec{\ell}).
\end{align}
from the $\langle[\Delta\psi(\vec{\ell})]^2\rangle$ values estimated for the observations.

For the submillimetre observations, the de-biasing correction is given by Equation~B.4. from \cite{houde2009}, which corresponds to 
\begin{equation}\label{eq:CorrSubmm}
C^{2}_{\rm submm} = \sigma^2_\psi(\vec{x}) + \sigma^2_\psi(\vec{x}+\vec{\ell}) - 2 \sigma_\psi(\vec{x})\sigma_\psi(\vec{x}+\vec{\ell})\,e^{-\ell^{2}/4W^{2}}
\end{equation}
where $W=0.425$\,FWHM is the beam radius.

\begin{table*}  % table* is a two-column table.  Drop the * for one column.
\begingroup
\newdimen\tblskip \tblskip=5pt
\caption{Quantities derived from the angular structure function $S_{2}(\ell)$.}
\label{table-sf}                            % Label goes here.
\nointerlineskip
\vskip -3mm
\footnotesize
\setbox\tablebox=\vbox{
   \newdimen\digitwidth 
   \setbox0=\hbox{\rm 0} 
   \digitwidth=\wd0 
   \catcode`*=\active 
   \def*{\kern\digitwidth}
   \newdimen\signwidth 
   \setbox0=\hbox{+} 
   \signwidth=\wd0 
   \catcode`!=\active 
   \def!{\kern\signwidth}
\halign{\hbox to 1.15in{#\leaderfil}\tabskip 2.2em&
\hfil#&\hfil#&\hfil#&\hfil#&\hfil#&\hfil#\tabskip 0pt\cr
\noalign{\doubleline}
%\omit\hfil Region\hfil & \multispan2\hfil At $\ell <10\arcmin$\hfil & \multispan1\hfil At $\ell >10\arcmin$\hfil \cr
\omit\hfil Region\hfil  & \hfil $\Delta S_{2}(1.5\arcmin)$$^{a,b}$ \hfil & \hfil $\Delta S_{2}(10.2\arcmin)$$^{a,c}$ \hfil & \hfil $\langle\Delta S_{2}(\ell)\rangle_{\ell >10\arcmin}$ \hfil \cr
%\omit\hfil Region\hfil & \hfil $<\varsigma_{\psi_{\rm star}}>$ \hfil &  \hfil $\tilde{\varsigma}_{\psi_{\rm star}}$$^a$  \hfil & \hfil$<\bar{\psi}_{\rm star}-\bar{\psi}_{\rm submm}>$  \hfil & \hfil  max$(\bar{\psi}_{\rm star}-\bar{\psi}_{\rm submm})$  \hfil & \hfil $S^{\rm star}_{2}(10\arcmin)$ \hfil & \hfil $S^{\rm star}_{2}(10\arcmin)-S^{\rm submm}_{2}(10\arcmin)$ \hfil \cr
\omit & [deg]\hfil & [deg]\hfil & [deg]\hfil \cr
\noalign{\vskip 4pt\hrule\vskip 6pt}
%----------------------------------------------------------------------------------------------------------------
%Taurus & \hfil10.4\hfil & \hfil\phantom{0}9.5\hfil & \hfil-0.2\hfil & \hfil28.5\hfil & \hfil18.2\hfil & \hfil\phantom{0}9.1\hfil \cr
Taurus & \hfil\phantom{0}7.6\hfil & \hfil14.5\hfil & \hfil $-$$^{d}$ \hfil \cr
%----------------------------------------------------------------------------------------------------------------
%Pipe & \hfil\phantom{0}7.5\hfil & \hfil\phantom{0}6.0\hfil & \hfil\phantom{0}1.1\hfil & \hfil\phantom{0}9.1\hfil & \hfil11.1\hfil & \hfil\phantom{0}3.9\hfil\cr
Pipe & \hfil\phantom{0}7.3\hfil & \hfil\phantom{0}9.3\hfil & \hfil\phantom{0}7.8$\,\pm\,$2.5\hfil \cr
%----------------------------------------------------------------------------------------------------------------
Lupus I & \hfil12.5\hfil & \hfil20.7\hfil & \hfil14.7$\,\pm\,$4.4\hfil  \cr
%----------------------------------------------------------------------------------------------------------------
%Musca & \hfil\phantom{0}6.2\hfil & \hfil\phantom{0}6.1\hfil & \hfil\phantom{0}1.5\hfil & \hfil\phantom{0}5.1\hfil & \hfil\phantom{0}9.8\hfil & \hfil\phantom{0}5.9\hfil\cr
Musca & \hfil\phantom{0}6.0\hfil & \hfil\phantom{0}8.4\hfil & \hfil\phantom{0}7.7$\,\pm\,$0.4\hfil \cr
%----------------------------------------------------------------------------------------------------------------
\noalign{\vskip 3pt\hrule\vskip 4pt}}}
\endPlancktablewide                    % ends one-column \halign
\tablenote a $\Delta S_{2}(\ell)$ as defined in Eq.~\ref{eqn:deltaS2}.\par %First distance bin.\par
\tablenote b 1.5\arcmin\ corresponds to the centre of the first $\ell$-bin in the range $0\arcmin<\ell<2.9\arcmin$.\par
\tablenote c 10.2\arcmin\ corresponds to the centre of the $\ell$-bin in the range $8.8\arcmin<\ell<11.7\arcmin$.\par
\tablenote d Towards Taurus, we find \stwostar\,$<$\,\stwosubmm\ for a wide range of $\ell$. This effect, produced by an insufficient amount of starlight polarization observations to cover those scales, makes the corresponding values of $\langle\Delta S_{2}(\ell)\rangle_{\ell >10\arcmin}$ imaginary and not meaningful.\par
\endgroup
\end{table*}  

% ===========================================================================================================
% DISCUSSION DISCUSSION DISCUSSION
% ===========================================================================================================
\section{Discussion}\label{section:discussion}

We use two methods to compare the \bperp\ structure as sampled by starlight and submillimetre, whose main difference is the range of scales over which the comparison is made.

In the vicinity method, we quantify the dispersion of \bperp\ orientation within the \Planck\ beam, which is accessible through the high angular resolution of the starlight polarization observations. 
The difference between \bperpplanck\ and \bperpstars\ averaged over the area of the \Planck\ beam amounts to a comparison at the 10\arcmin\ scale.
To relate this difference to the structure of the field at smaller scales, we compare it with $\varsigma_{\psi_{\rm star}}$ and $\varsigma_{p_{\rm star}}/\bar{p}_{\rm star}$.

In the \stwo\ method, we consider the dispersion of the \bperpplanck\ and \bperpstars\ orientation angles across multiple scales. 
At scales below the size of the \Planck\ beam, $\ell < 10$\arcmin, we evaluate how much structure is smoothed by the angular resolution of the \Planck\ 353\,GHz observation. 
At scales above the size of the \Planck\ beam, $\ell >10$\arcmin, we evaluate if the structure of the field traced by starlight polarization is comparable to that inferred from the  \Planck\ 353\,GHz observations.

\subsection{The \bperp\ structure within the \Planck\ beam}

The average values of the difference between \bperpstars\ and \bperpplanck\ within the 10\arcmin-diameter vicinities, presented in Table~\ref{table-means} and illustrated in Fig.~\ref{fig:StarAnglevsSubmmAngle}, show that the \bperp\ orientations inferred from optical/NIR polarization angle follow those inferred from submillimetre observations within approximately 5\degr, thus confirming the visual impression from Fig.~\ref{fig:StarlightandLICsubmm} and suggesting excellent agreement between the \bperp\ orientations estimated with both techniques when evaluated at the 10\arcmin\ scale.

The values in Table~\ref{table-means} reveal that the 85-th and 95-th percentiles of $|\bar{\psi}_{\rm star}-\bar{\psi}_{\rm submm}|$ in the vicinities are less than 12\deg\ and 16\pdeg6, respectively.
The left panel of Fig.~\ref{fig:DiffVsSigmaPsi} shows that the distribution of the values $|\bar{\psi}_{\rm star}-\bar{\psi}_{\rm submm}|$ and $\varsigma_{\psi_{\rm star}}$ is different in each region.
Also, there seems to be a trend in $|\bar{\psi}_{\rm star}-\bar{\psi}_{\rm submm}|$ and $\varsigma_{\psi_{\rm star}}$, as the largest differences between the orientation angles correspond the vicinities with the largest dispersions of orientation angles.
Towards the Musca region, all the vicinities are grouped around $|\bar{\psi}_{\rm star}-\bar{\psi}_{\rm submm}| \lesssim 5\deg$ and $\varsigma_{\psi_{\rm star}} \lesssim9\deg$.
Towards the Pipe Nebula, we find that the majority of the vicinities have $|\bar{\psi}_{\rm star}-\bar{\psi}_{\rm submm}| \lesssim 15\deg$, but there are many vicinities with $\varsigma_{\psi_{\rm star}} \gtrsim30$\deg. 
Towards the Lupus I region, where the mean values of $|\bar{\psi}_{\rm star}-\bar{\psi}_{\rm submm}|$ and $\varsigma_{\psi_{\rm star}}$ are the largest, there are many vicinities with $\varsigma_{\psi_{\rm star}} \gtrsim40$\deg.

To estimate the amount of dispersion that would be expected just from the differences in angular resolution between the two observations, we apply the same analysis to the Gaussian polarization models introduced in Sect.~\ref{section:GaussianRealization}.
The behaviour of $|\bar{\psi}_{2}-\bar{\psi}_{10}|$ with respect to $\varsigma_{\psi_{2}}$ is only related to the difference in the angular resolutions since, by construction, $\bar{\psi}_{10}$ and $\bar{\psi}_{2}$ correspond to the same field. %even when we observe large values of $\varsigma_{\psi_{2}}$.

The right panel of Fig.~\ref{fig:DiffVsSigmaPsi} shows that these simple models qualitatively reproduce some of the trends seen towards the considered MCs.
In this example, the models with $f_{\rm M}=0.5$ reproduce the low mean and dispersion of $|\bar{\psi}_{\rm star}-\bar{\psi}_{\rm submm}|$ and $\varsigma_{\psi_{\rm star}}$ values seen towards the Musca regions.
The models with larger dispersion around the mean field, $f_{\rm M}=1.0$, show similar trends to the Lupus I and the Pipe Nebula, where the values of $\varsigma_{\psi_{\rm star}}$ are larger than in Musca.
However, one should refrain from drawing conclusions on the values of $\alpha_{\rm M}$ and $f_{\rm M}$ towards these regions just from this comparison as these parameters are degenerate and the inclination of the mean field orientation with respect to the plane of the sky is unknown.
What can be learned from the comparison with these simple polarization models is that the observed values of $|\bar{\psi}_{\rm star}-\bar{\psi}_{\rm submm}|$ and $\varsigma_{\psi_{\rm star}}$ do not indicate that stars and the submillimetre observations are sampling significantly different \bvec\ structures.

To further investigate the relation between $|\bar{\psi}_{\rm star}-\bar{\psi}_{\rm submm}|$ and the structure within the \Planck\ beam, we compare this quantity with the fluctuations in the starlight polarization fraction characterized by $\varsigma_{p_{\rm star}}/\bar{p}_{\rm star}$.
In principle, each $\bar{p}_{\rm star}$ observation carries information about the dispersion of the field along the LOS that might be lost if we only consider the orientation angle.

The left panel of Fig.~\ref{fig:DiffVsSigmaPoverP} shows that for most of the vicinities the fluctuations of $p_{star}$ are relatively low, $\varsigma_{p_{\rm star}}/\bar{p}_{\rm star} < 1$. 
The largest values of $\varsigma_{p_{\rm star}}/\bar{p}_{\rm star}$, present mainly towards Lupus I and the Pipe Nebula, are not particularly associated with the largest values of $|\bar{\psi}_{\rm star}-\bar{\psi}_{\rm submm}|$.
It is tempting to interpret large fluctuations of $p_{\rm star}$ within a vicinity as indicating that each star is sampling considerably different media along the LOS. 
However, the results of the analysis of the Gaussian models, presented in the right panel of Fig.~\ref{fig:DiffVsSigmaPoverP}, indicate that these fluctuations can be produced by the dispersions in the 3-dimensional orientation of \bvec, which are larger for larger values of $f_{\rm M}$ or values of $\alpha_{\rm M}$ closer to zero. 
The similarities in the behaviour of $|\bar{\psi}_{2}-\bar{\psi}_{10}|$ with respect to $\varsigma_{2}/\bar{p}_{2}$ and $|\bar{\psi}_{\rm star}-\bar{\psi}_{\rm submm}|$ with respect to $\varsigma_{p_{\rm star}}/\bar{p}_{\rm star}$ do not indicate that starlight and submillimetre observations are sampling significantly different \bvec\ structures.
But they indicate considerable differences in the dispersion of \bvec\ in the different regions.

In principle, the value of the dispersions $\varsigma_{p_{\rm star}}/\bar{p}_{\rm star}$ and $\varsigma_{\psi_{\rm star}}$ could be associated with the amount of turbulence in each region. 
However, the velocity dispersions in each region observed at angular resolutions close to 10\arcmin\ are not significantly different, with $\sigma_{v}=1.2\pm0.5$, $1.5\pm0.6$, and $1.0\pm0.4$\,km\,s$^{-1}$ towards Taurus, Lupus, and the Chamaeleon-Musca regions, respectively\footnote{These values are estimated from CO emission-line observations \citep{dame2001} and presented in Table~D.1. of \cite{planck2015-XXXV}}. 
But these MCs are located in particularly different environments.
Recent studies indicate that Lupus I, the region where we see the largest values of $\varsigma_{p_{\rm star}}/\bar{p}_{\rm star}$ and $\varsigma_{\psi_{\rm star}}$, has a larger star formation rate and its formation is associated with large feedback events \citep{rygl2013,gaczkowski2015}.
In contrast, Musca and Taurus, where the values of $\varsigma_{p_{\rm star}}/\bar{p}_{\rm star}$ and $\varsigma_{\psi_{\rm star}}$ are low, are apparently more quiescent \citep{kenyon2008,luhman2008}.
Nevertheless, establishing a detailed relation between the gas kinematics in each of these clouds and the structure of \bperp\ is not straightforward and it is beyond the goal of this work.

For the sake of completeness, we evaluate the relation between $\varsigma_{p_{\rm star}}/\bar{p}_{\rm star}$ and $\varsigma_{\psi_{\rm star}}$.
Fig.~\ref{fig:SigmapOverMeanpVsSigmaPsi} shows how the values of $\varsigma_{\psi_{\rm star}}$ are well correlated with $\varsigma_{p_{\rm star}}/\bar{p}_{\rm star}$.
However, the values of $\varsigma_{p_{\rm star}}/\bar{p}_{\rm star}$ show a larger dynamic range since the angle dispersion saturates at $\varsigma_{\psi_{\rm star}}\approx 52\deg$, while $\varsigma_{p_{\rm star}}$ in unbounded.

The trend in the observed values of $\varsigma_{p_{\rm star}}/\bar{p}_{\rm star}$ and $\varsigma_{\psi_{\rm star}}$ in the vicinities is consistent with the results found in the Gaussian models.
%, particularly those with a large amount of small scale structure, $\alpha_{\rm M}=-1.5$.
This is important because such a trend is not possible to reproduce in terms of the dust grain alignment efficiency alone.
In the observations, we find that the largest values of $\varsigma_{p_{\rm star}}/\bar{p}_{\rm star}$ are consistently associated with the largest $\varsigma_{\psi_{\rm star}}$.
If the dust grain alignment was the only process responsible for the observed \bperpstars, the decrease in the values of $p_{\rm star}$ would be independent of the values of $\varsigma_{\psi_{\rm star}}$ since the lowest values of $p_{\rm star}$ would correspond to vicinities where the dust polarization does not sample the \bvec\ morphology making the observed \bperpstars\ orientations random.
This accounts for the saturation of $\varsigma_{\psi_{\rm star}}\approx 52\deg$, but it does not account for the correlation between $\varsigma_{p_{\rm star}}/\bar{p}_{\rm star}$ and $\varsigma_{\psi_{\rm star}}$ at smaller $\varsigma_{\psi_{\rm star}}$ values.
This does not exclude the effect of the dust grain alignment efficiency, which we have assumed to be perfect, but indicates that the magnetic field morphology and the spatial correlations in the polarization observations are a key ingredient for the interpretation of the distributions of $p$ and $\psi$.
  
\begin{figure*}[ht!]
\centerline{
\includegraphics[width=0.47\textwidth,angle=0,origin=c]{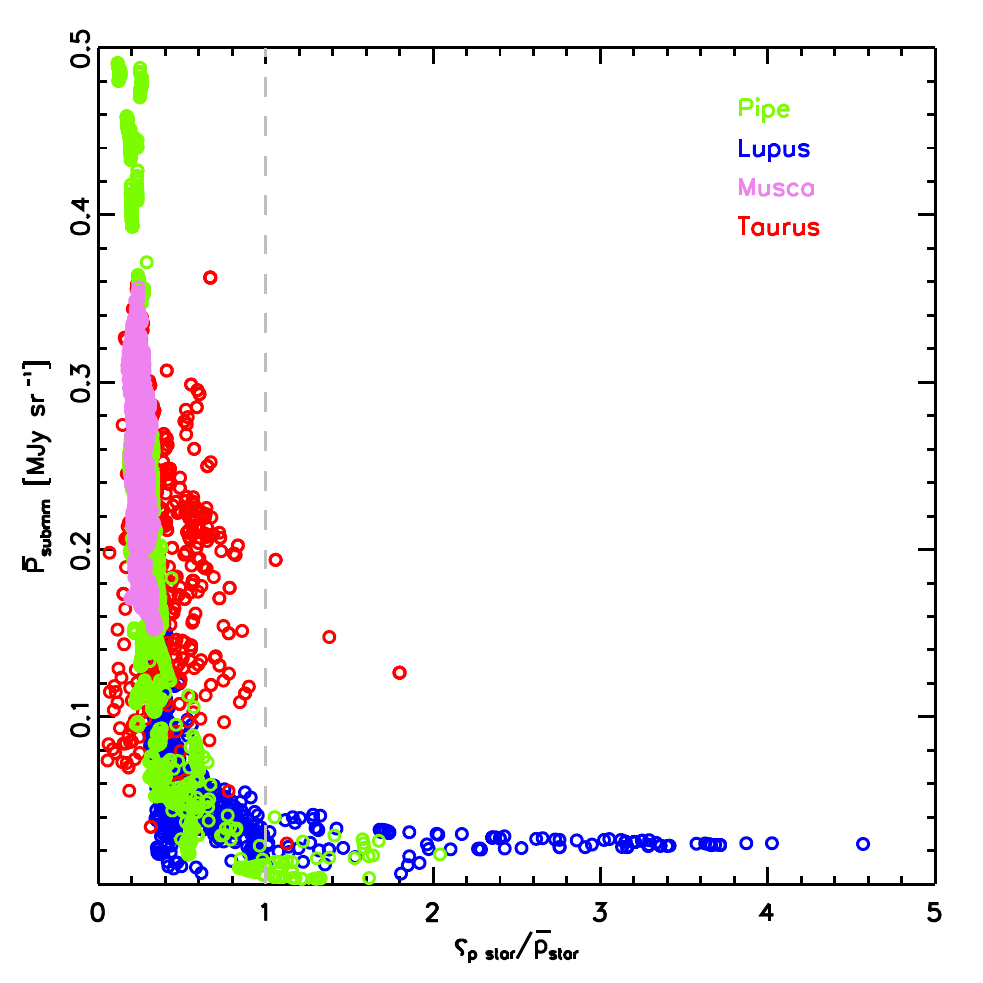}
\hspace{-0.5cm}
\includegraphics[width=0.47\textwidth,angle=0,origin=c]{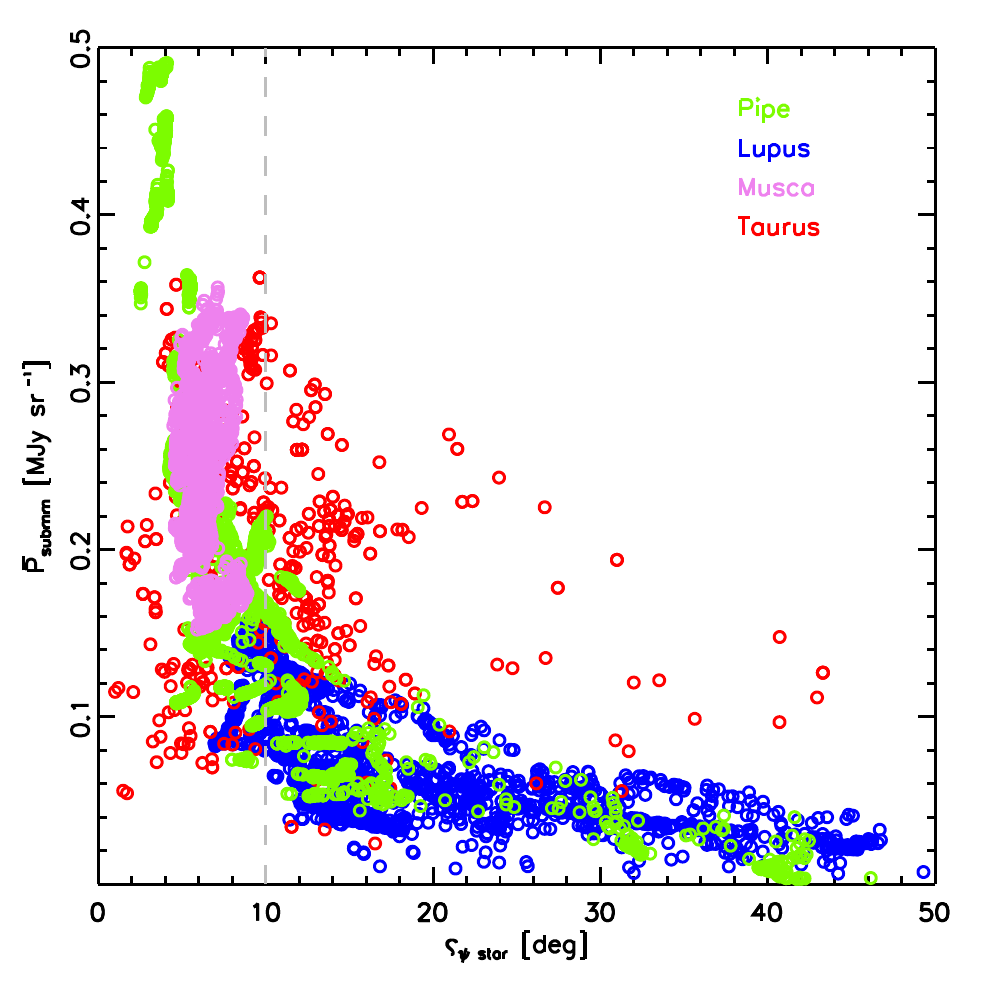}
}
\caption{Scatter plot of the fluctuations in polarized fraction, $\varsigma_{p_{\rm star}}/\bar{p}_{\rm star}$ (\emph{left}), and the dispersion of orientation angles, $\varsigma_{\psi_{\rm star}}$ (\emph{right}), against the mean values of the total polarized flux from the \Planck\ 353\,GHz observations in all the 10\arcmin\ vicinities with more than three stars towards each observed region. The dashed grey lines, included just for reference, correspond to $\varsigma_{p_{\rm star}}/\bar{p}_{\rm star}=1$ (\emph{left}) and $\varsigma_{\psi_{\rm star}}=10$\deg\ (\emph{right}).}
\label{fig:MeanP353VsSigmapOverMeanp}
\end{figure*}

\begin{figure}[ht!]
\centerline{
\hspace{-0.5cm}
\includegraphics[width=0.45\textwidth,angle=0,origin=c]{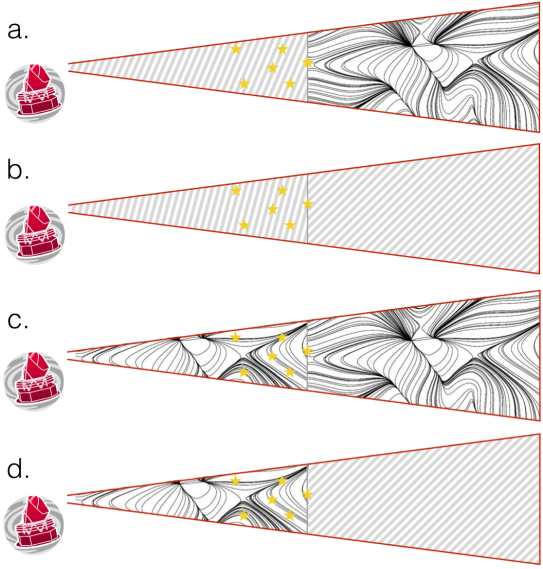}
}
\caption{Illustration of four simplified models of the LOS sampled by a \Planck\ beam; (a) stars  located in a portion of the ISM where \bvec\ is uniform and beyond the last star \bvec\ is not uniform, (b) stars located in a portion of the ISM where \bvec\ is uniform and beyond the last star \bvec\ is uniform with a possibly different mean direction, (c) stars located in a portion of the ISM where \bvec\ is not uniform and beyond the last star \bvec\ is not uniform, and (d) stars located in a portion of the ISM where \bvec\ is not uniform and beyond the last star \bvec\ is uniform.}
\label{fig:SchemaStars}
\end{figure}

\begin{figure*}[ht!]
\centerline{
\includegraphics[width=0.48\textwidth,angle=0,origin=c]{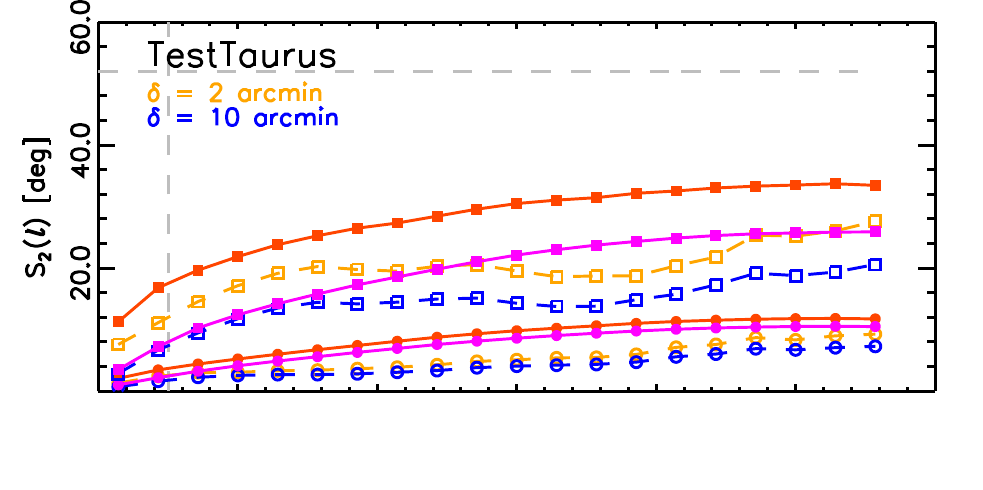}
\includegraphics[width=0.48\textwidth,angle=0,origin=c]{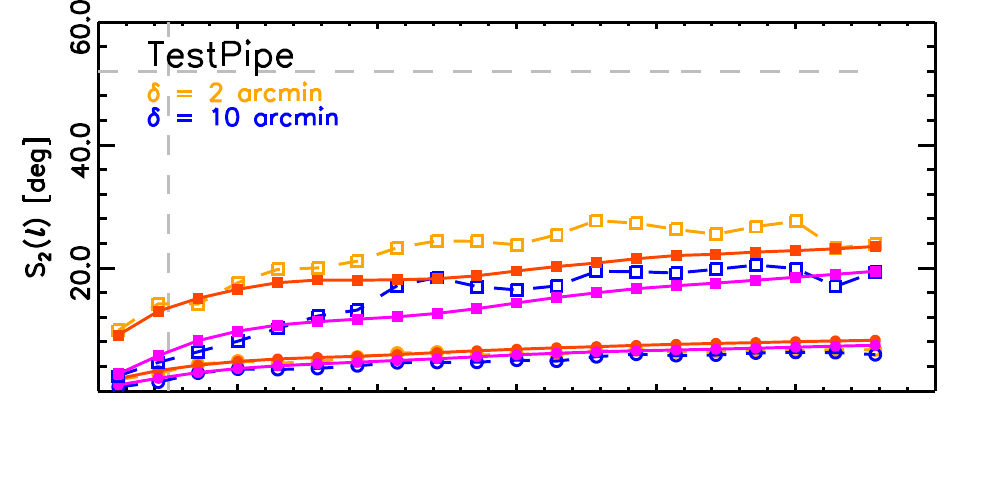}
}
\vspace{-0.95cm}
\centerline{
\includegraphics[width=0.48\textwidth,angle=0,origin=c]{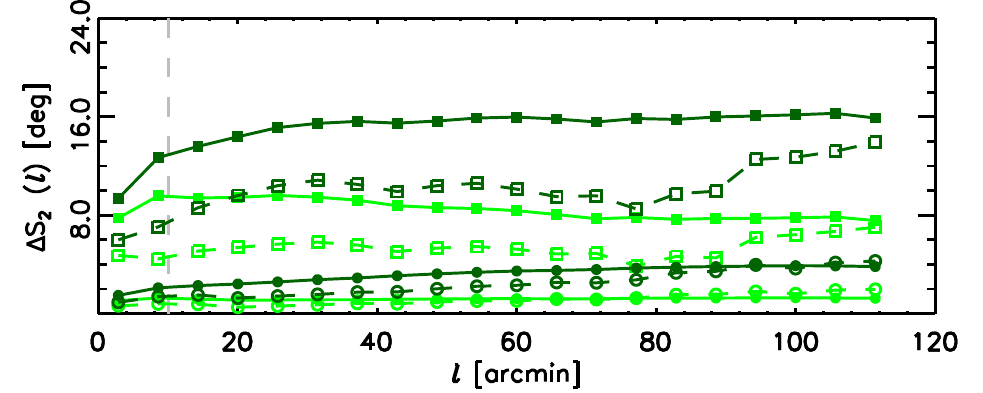}
\includegraphics[width=0.48\textwidth,angle=0,origin=c]{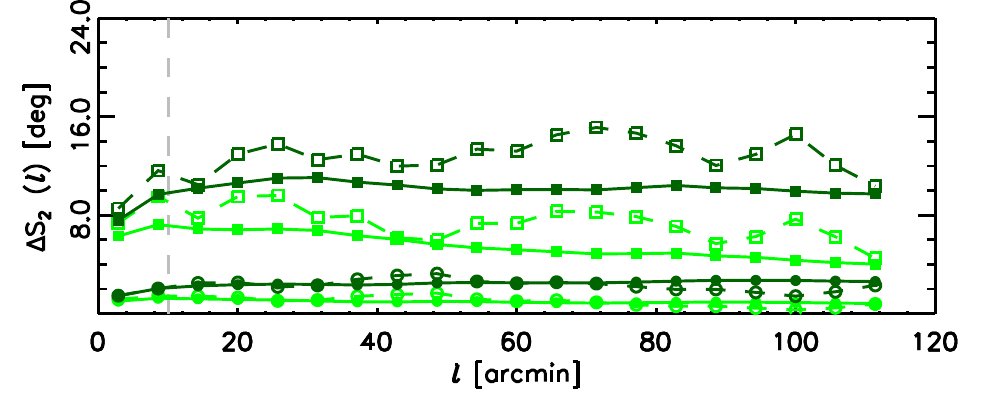}
}
\vspace{-0.1cm}
\centerline{
\includegraphics[width=0.48\textwidth,angle=0,origin=c]{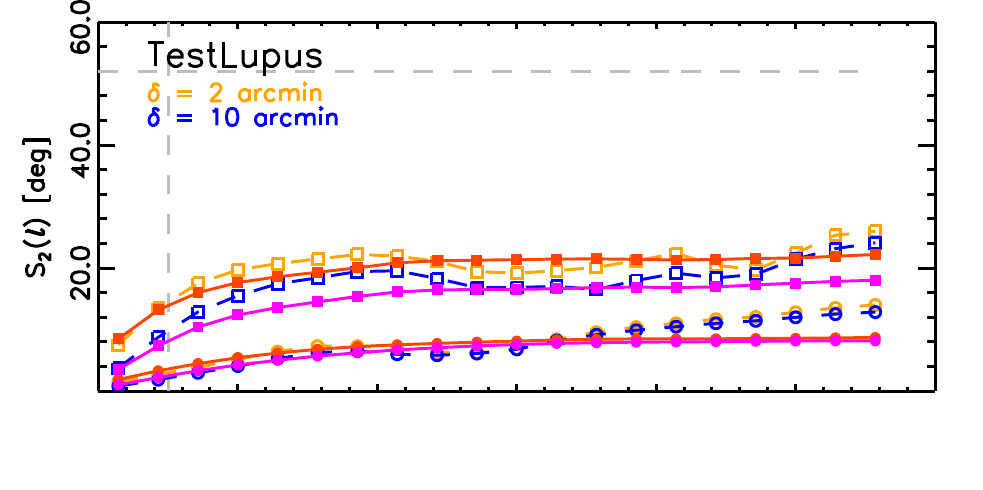}
\includegraphics[width=0.48\textwidth,angle=0,origin=c]{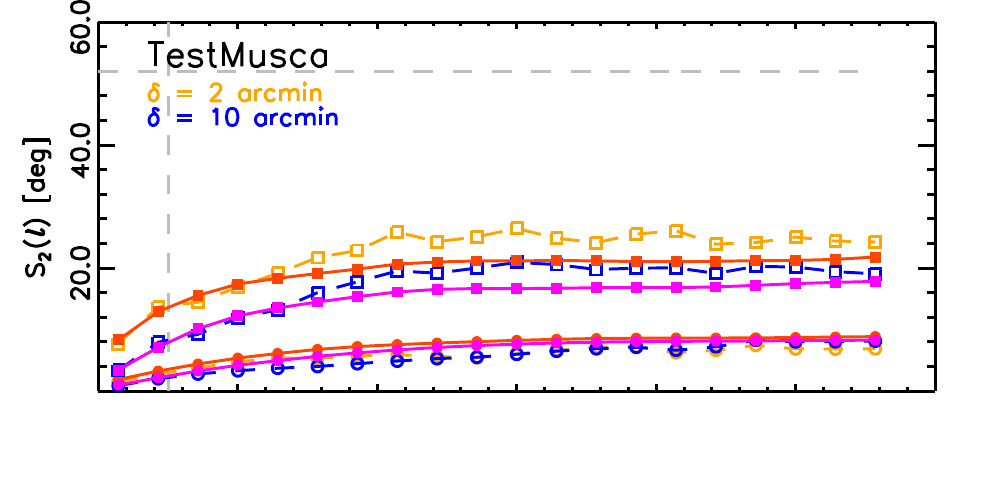}
}
\vspace{-0.95cm}
\centerline{
\includegraphics[width=0.48\textwidth,angle=0,origin=c]{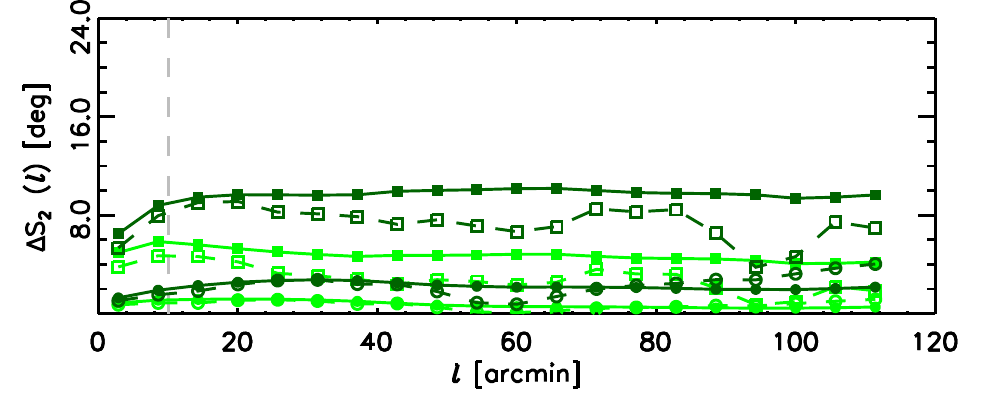}
\includegraphics[width=0.48\textwidth,angle=0,origin=c]{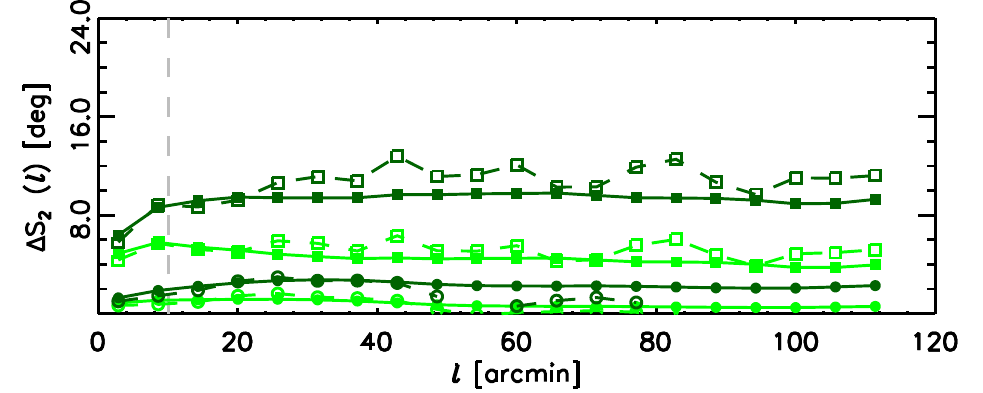}
}
\caption{
Structure functions \stwo\ calculated from the Gaussian polarization models corresponding to $f_{\rm M} =1.0$  and $\alpha_{\rm M}=-2.5$ (circles) and $\alpha_{\rm M}=-1.5$ (squares), with angular resolutions $\delta = 2$\arcmin\ (orange and red), 10\arcmin\ (blue and magenta). 
The dashed lines with open symbols, shown in orange and blue, correspond to \stwo\ calculated using only the values of $Q_{\rm M}$ and $U_{\rm M}$ towards lines of sights with starlight polarization observations towards each region.
The solid lines with filled symbols, shown in red and magenta, correspond to \stwo\ calculated using 10,000 LOSs randomly distributed over each region.
In the lower plot of each panel, we present the differences,
$\Delta S_{2}(\ell) \equiv \left([S^{\rm \delta=2'}_{2}(\ell)]^{2}-[S^{\rm \delta=10'}_{2}(\ell)]^{2}\right)^{1/2}$ (dark green), and $\delta S_{2}(\ell) \equiv S^{\rm \delta=2'}_{2}(\ell)-S^{\rm \delta=10'}_{2}(\ell)$ (light green).
There is no noise included in the Gaussian models of polarization, so no bias correction is necessary.
}
\label{fig:StructureFunctionModels}
\end{figure*}

\subsection{Background and line of sight depth}

So far we have interpreted the values of $|\bar{\psi}_{\rm star}-\bar{\psi}_{\rm submm}|$, $\varsigma_{\psi_{\rm star}}$, and $\varsigma_{p_{\rm star}}/\bar{p}_{star}$ in terms of Gaussian models of polarization, which are 3-dimensional in their treatment of \bvec, but do not include the depth of the LOS, which is different for the starlight and submillimetre observations.

The magnetic field responsible for the starlight polarization angle observed towards a particular direction is the average of various components encountered by the beam of starlight as it traverses the ISM, weighted by the specific extinction in each segment of path.
In contrast, the \Planck\ observations include the contributions of the dust thermal polarized emission along all the LOS.
If we assume that the dust scale height is 50\,pc and the mean density is $n_{0}\approx 1$\,cm$^{-3}$, the total gas column density contributed by the dust in the Galactic disk corresponds to about 40\% of the mean \nh\ towards the vicinities in the considered regions.
Given that the stars within the same \Planck-beam-sized vicinity can be distributed at multiple distances, thus sampling different segments in the ISM, we can potentially use the starlight polarization observations to characterize \bperp\ in different portions of the LOS and describe its structure using the observed values of $\varsigma_{p_{\rm star}}/\bar{p}_{\rm star}$ and $\varsigma_{\psi_{\rm star}}$.
%The stars within the same \Planck-beam-sized vicinity can be distributed at multiple distances, thus sampling different segments in the ISM. 
%Those differences in LOS depth can potentially be responsible for the observed values of $\varsigma_{p_{\rm star}}/\bar{p}_{\rm star}$ and $\varsigma_{\psi_{\rm star}}$.

Unfortunately, the distance to the majority of the stars included in this study is currently unknown, making it difficult to assess the contribution of each section of the LOS to the total polarized signal.
For the moment, we can compare the behaviour of $\varsigma_{\psi_{\rm star}}$ and $\varsigma_{p_{\rm star}}/\bar{p}_{\rm star}$ as a function of the total polarized flux in the \Planck\ 353\,GHz observations and characterize any correlations that indicate the differences in the LOS depth and the potential effect of an homogeneous polarization background behind the stars.

Fig.~\ref{fig:MeanP353VsSigmapOverMeanp} presents a comparison between the $\varsigma_{p_{\rm star}}/\bar{p}_{\rm star}$ and $\varsigma_{\psi_{\rm star}}$ with the mean values of the total polarized flux from the \Planck\ 353\,GHz observations, $\bar{P}_{\rm submm}$, within the 10\arcmin\ vicinities.
We observe that the largest values of $\varsigma_{p_{\rm star}}/\bar{p}_{\rm star}$ and $\varsigma_{\psi_{\rm star}}$ are in general associated with the lowest values of $\bar{P}_{\rm submm}$.
We do not find vicinities where $\bar{P}_{\rm submm}$ is large and where, simultaneously, the values of $\varsigma_{p_{\rm star}}/\bar{p}_{\rm star}$ and $\varsigma_{\psi_{\rm star}}$ are also large corresponding to a large dispersion of \bperp\ or large differences in the LOS sampled by different stars within the same vicinity.
We interpret this as an indication that there is no contribution of an homogeneous polarized background behind the stars.

To better describe this interpretation, we illustrate our argument with four toy models of the possible \bperp\ structure sampled by the \Planck\ beam along the LOS, shown in Fig.~\ref{fig:SchemaStars}. We consider that:
\begin{enumerate}[(a)]%for capital roman numbers.
\item the stars are located in a portion of the ISM where \bvec\ is uniform and beyond the last star \bvec\ is not uniform, 
\item the stars are located in a portion of the ISM where \bvec\ is uniform and beyond the last star \bvec\ is uniform with a possibly different mean direction, 
\item the stars are located in a portion of the ISM where \bvec\ is not uniform and beyond the last star \bvec\ is not uniform,
\item the stars are located in a portion of the ISM where \bvec\ is not uniform and beyond the last star \bvec\ is uniform.
\end{enumerate}

For model (a), we expect relatively low values of $\varsigma_{p_{\rm star}}/\bar{p}_{\rm star}$ and $\varsigma_{\psi_{\rm star}}$, since the uniform structure of the field introduces correlations between the polarization of stars at the different distances; relatively low values of $|\bar{\psi}_{\rm star}-\bar{\psi}_{\rm submm}|$, since the projected field orientation is dominated by the portion of the ISM where \bperp\ is homogeneous; and not particularly low values of $\bar{P}_{\rm submm}$.

For model (b), we expect relatively low values of $\varsigma_{p_{\rm star}}/\bar{p}_{\rm star}$ and $\varsigma_{\psi_{\rm star}}$, for the same reason mentioned in the previous case; homogeneous values of $|\bar{\psi}_{\rm star}-\bar{\psi}_{\rm submm}|$ that correspond to the difference between the mean \bperp\ in the first portion and the average \bperp\ from both portions of the LOS; and not particularly low values of $\bar{P}_{\rm submm}$, unless the two regions happen to have fields at 90\deg\ to each other.

For model (c), we expect relatively large values of $\varsigma_{p_{\rm star}}/\bar{p}_{\rm star}$ and $\varsigma_{\psi_{\rm star}}$, given that the correlation length of \bperp\ is small; values of $|\bar{\psi}_{\rm star}-\bar{\psi}_{\rm submm}|$ that depend on the length of each portion of the LOS; and low values of $\bar{P}_{\rm submm}$.

For model (d), we expect relatively large values of $\varsigma_{p_{\rm star}}/\bar{p}_{\rm star}$ and $\varsigma_{\psi_{\rm star}}$, for the same reason mentioned in the previous case; values of $|\bar{\psi}_{\rm star}-\bar{\psi}_{\rm submm}|$ that depend on the structure of the portion of the LOS portion sampled by the stars; and not particularly low values of $\bar{P}_{\rm submm}$, which are dominated by the \bvec\ correlation in the medium behind the stars.

The behaviour of the Lupus I and portions of the Pipe Nebula seems to be related to model (c). 
In contrast, the behaviour of Musca is more similar to model (b), as previously identified in \cite{planck2014-XXXIII}, where the estimation of the `background'' was inferred using the polarization orientation at different \nh. 
Model (a) is harder to evaluate, given that we do not have a clear estimate of how much of $\bar{P}_{\rm submm}$ is due to the portion of the LOS also sampled by the stars, but the results of the analysis do not discard it. What it is clear from this test is that there is no evidence of model (d).
The stars considered in this study can be located as far as 1 or 2\,kpc from the Sun, thus it is not possible to unambiguously define what is the contribution of the ISM behind the cloud to the observed \bperpplanck.

To further advance in the investigation of the polarized background, one could compare the values of $P_{\rm submm}$ and $p_{\rm star}$ towards high Galactic latitudes, where the depth of the LOS is limited and the contribution of the background is negligible. 
However, there is evidence that the dust towards the aforementioned regions has different properties than the dust in the MCs \citep{planck2014-XXIX} and this involves a detailed study of the dust emission and extinction properties, which is beyond the scope of this work.

% --------------------------------------------------------------------------------------------------------------------------------------------------------------------------------------------------------------------
\subsection{The angular structure function \stwo}

Our objective in computing \stwostar\ and \stwosubmm\ along the same LOSs, shown in Fig.~\ref{fig:StructureFunctionZoom}, is evaluating the differences between the \bperpstars\ and \bperpplanck\ structure across multiple scales.
The differences $\delta S_{2}(\ell)$ and $\Delta S_{2}(\ell)$, defined in Eq.~\ref{eq:deltaS2lin} and Eq.~\ref{eqn:deltaS2} respectively, are shown in the lower plots of each panel in Fig.~\ref{fig:StructureFunctionZoom} and summarized in Table.~\ref{table-sf}.
We also test if the behaviour of \stwostar\ and \stwosubmm\ is representative of the \bperp\ behaviour over each region by comparing it with \stwoallsubmm, which corresponds to 10,000 LOSs randomly distributed over the \Planck\ 353\,GHz polarization maps.
Given the scale set by the angular resolution of the \Planck\ observations, we discuss separately the behaviour of $S_{2}(\ell)$ at scales $\ell < 10$\arcmin\ and $\ell > 10$\arcmin.

\subsubsection{The structure function at $\ell < 10$\arcmin}

At scales $\ell < 10$\arcmin, \stwosubmm\ progressively tends to zero with decreasing $\ell$ as expected from the smoothing by the \Planck\ beam.
In the lowest $\ell$-bin, which corresponds to the range $0\arcmin<\ell<2.9\arcmin$, we find that the differences between \stwostar\ and \stwosubmm\ lie between 6\pdeg0 and 12\pdeg5, as presented in Table~\ref{table-sf}.
The values of the bias corrections, $C_{\rm star}$ and $C_{\rm submm}$ shown in the lower plot of each panel in Fig.~\ref{fig:StructureFunctionZoom}, indicate that these differences are not due to noise, but correspond to the structure of \bperp\ at $\ell<2.9\arcmin$, which has been characterized in previous studies \citep{franco2010,chapman2011,franco2015}.

In the $\ell$-bin around $\ell=10\arcmin$, which corresponds to the range $8.8\arcmin<\ell<11.7\arcmin$, we find that the differences between \stwostar\ and \stwosubmm\ lie between 8\pdeg5 and 20\pdeg7, as also presented in Table~\ref{table-sf}.
These values are directly comparable to those presented in Table~\ref{table-means}, but given that the quadratic averaging implied in the calculation of \stwo\ enhances large dispersion values, it is expected that $\Delta S_{2}(\ell\approx10\arcmin) > \langle\varsigma_{\psi_{\rm star}}\rangle$.
The largest values of $\Delta S_{2}$ at $\ell < 10$\arcmin\ are found towards the Lupus I region, where the vicinity analysis also identifies the largest values of $\langle\varsigma_{\psi_{\rm star}}\rangle$.
Consistently, the Pipe Nebula and the Musca regions have the lowest values of $\langle\varsigma_{\psi_{\rm star}}\rangle$ and $\Delta S_{2}$ at $\ell < 10$\arcmin.

\subsubsection{The structure function at $\ell > 10$\arcmin}

At scales $\ell > 10$\arcmin, the values of \stwostar\ and \stwosubmm\ are roughly constant, but present wave-like features or ``jitter''. 
The ``jitter'' is not present in \stwoallsubmm, which is estimated using $Q_{\rm submm}$ and $U_{\rm submm}$ in 10,000 LOSs randomly distributed over each region and not just along the lines of sight with observations of starlight polarization, thus suggesting that these features are related to the distribution of the observed stars in the plane of the sky and not to the structure of \bperp. 
To evaluate this LOS distribution effect, we compute \stwo\ in the Gaussian polarizations models introduced in Sect.~\ref{section:GaussianRealization}. 

For illustration purposes, we present in Fig.~\ref{fig:StructureFunctionModels} the values of \stwo\ computed with the Stokes parameters $Q_{\rm M}$ and $U_{\rm M}$ corresponding to the models with either $\alpha_{M}=-1.5$ and $\alpha_{M}=-2.5$, $f_{\rm M}=1.0$ and $\gamma=20$\deg. 
The results of this simple test reveal that the distribution of the stars introduces ``jitter'' features in \stwo\ that depend on the parameters in the model, but that are largely attenuated when considering a large amount of LOSs uniformly distributed over each region.

The ``jitter'' is largest in the \stwo\ values corresponding to the model with the least amount of spatial correlation, $\alpha_{\rm M}=-1.5$.
In the model with $\alpha_{\rm M}=-2.5$, the effect of the sampling is less manifest, as larger regions of the polarization maps are correlated, but this effect is not the result of the spectral index of turbulence alone. 
Given that lower values of $f_{\rm M}$ values correspond to small dispersions around the mean field direction, they also correspond to larger spatial correlation than models with higher $f_{\rm M}$.
In the same manner, lower values of $\gamma$ correspond to mean field orientations closer to the plane of the sky, thus producing larger amount of spatial correlation than models where the mean field orientation is closer to the line of sight.

The ``jitter'' can be associated to two aspects of the spatial distribution of the stars in the plane of the sky. 
On the one hand, starlight polarization observations are grouped in fields that correspond to the telescope field of view, thus imposing a particular sampling scale, i.e., a set of distances that are much better sampled than others. 
This effect, which is less noticeable in the $S_{2}$ studies at scales below the size of the field of view \citep{franco2010,franco2015}, is difficult to identify as it would involve sparsely sampling the data, requiring a large amount of observations at multiple separations to populate the $\ell$-bins with enough statistical significance.
On the other hand, $S_{2}$ involves averaging over $\ell$ in all directions. This is not always possible with the stars, as the star separations are not necessarily isotropic and some of the $ell$-bins may be populated by stars distributed towards one particular direction.
Both of these effects do not affect the submillimetre observations, where all the scales of at least one half of the map size are isotropically sampled.

The observations of starlight and submillimetre polarization included in this study are not sufficient to constrain $\alpha_{\rm M}$, $f_{\rm M}$, and $\gamma$ towards the studied regions.
However, the behaviour of \stwo\ in the polarization models illustrates that the spatial distribution of the starlight polarization observations can significantly affect the values of \stwostar.
This is significant for the studies of the MC-scale magnetic field strengths calculated with the David-Chandrasekhar-Fermi method \citep[DCF,][]{davis1951a,chandrasekhar1953}.

\subsubsection{Dispersion of polarization angles and magnetic field strength}

In the DCF method, the calculated field strengths depend on the dispersion of the polarization angles, which is estimated directly using $\varsigma_{\psi_{\rm star}}$ \citep{chandrasekhar1953} or using \stwo\ \citep{hildebrand2009,houde2009}, the velocity dispersion, and the mean density in the considered region \citep[for a detailed description, see Appendix D of ][]{planck2015-XXXV}.
As observed in the vicinity and the \stwo\ analyses in the present work, the values of $\varsigma_{\psi_{\rm star}}$ and \stwo\ can be affected by the number and the distribution of starlight polarization observations.
Particularly towards the Pipe and the Taurus regions, it is evident that the considerable agreement between \stwostar\ and \stwosubmm, does not imply that the dispersion of the polarization angles is representative of the behaviour of \stwo\ obtained with a larger number of observations towards the same region, \stwoallsubmm.

Towards the Taurus region, \stwosubmm\ is unexpectedly larger than \stwostar\ in the range $50\arcmin < \ell < 90\arcmin$, thus producing $\delta S_{2}(\ell) < 0$ and imaginary values of $\Delta S_{2}(\ell)$.
This behaviour is due to the low number of observations in this range of separations, even after we included stars in the \cite{heiles2000} catalog to increase the sampling in the aforementioned $\ell$-range, hence, we do not further consider this region in the discussion of \stwo.

Towards the Pipe Nebula,  \stwostar\ and \stwosubmm\ show the presence of the ``jitter'' at multiple scales.
The values of $\Delta S_{2}(\ell)$ also show ``jitter'', including the two $\ell$-ranges around $\ell\approx90\arcmin$ and $\ell\approx110\arcmin$ where $\delta S_{2}(\ell) < 0$.
At $\ell>10\arcmin$, $\Delta S_{2}(\ell)$ is in average 7\pdeg8.

In this region and at $\ell > 10$\arcmin, \stwoallsubmm\ has a considerably different slope than \stwostar\ and \stwosubmm. 
This implies that \stwosubmm\ corresponds to a component of \bperp\ that is not dominant over most of the region, since the analysis of the polarization models indicates that the sampling of LOSs does not reproduce the differences between \stwosubmm\ and \stwoallsubmm.

If one considers only \stwostar\ and \stwosubmm, the angle dispersion term in the DCF method, which is directly related to the $y$-axis intercept of a Taylor series fit to the large scale component of the \stwo\ functions \citep{houde2009,houde2016}, the MC-scale magnetic field strength computed from the \Planck\ 353\,GHz polarization observations would be about 15\% larger than one inferred from the starlight polarization observations.
This estimate does not take into account the fact that the velocity dispersion may change when considering observations with higher angular resolution, but show that the increase in the angular resolution of polarization observations towards this region does not substantially change the results of the DCF method.

Towards Lupus I, both \stwostar\ and \stwosubmm\ present ``jitter'', but in contrast with the Pipe Nebula, \stwostar\ and \stwosubmm\ are close to \stwoallsubmm\ at $\ell > 0$\arcmin.
The average value of $\Delta S_{2}(\ell)$ at $\ell>10\arcmin$ is 14\pdeg7, with large variations at $\ell\approx80\arcmin$ and $\ell>100\arcmin$.
Following the same consideration described for the Pipe Nebula, the MC-scale magnetic field strength computed from the \Planck\ 353\,GHz polarization observations would be about 12\% larger than one inferred from the starlight polarization observations.

Towards Musca, \stwostar\ and \stwosubmm\ are less affected by the ``jitter'', as found in the polarization model with the largest amount of spatial correlations.
The values of \stwosubmm\ are close to \stwoallsubmm.
The values of $\Delta S_{2}(\ell)$ are also not significantly affected by the ``jitter'' and are in average 7\pdeg7.
In contrast with the aforementioned regions, the low average values of \stwostar\ and \stwosubmm\ at $\ell > 10$ produce a larger relative difference, about 70\%, between the field strengths estimated with the DCF method towards this region.

%If the sampling by the stars is sufficient, such that the effect of the ``jitter'' is negligible, the value of $\Delta S_{2}(\ell)$ constitutes the contribution of the small scale structure to the estimates of \bperp\ orientation dispersion used for computing the magnetic field strengths in the $S_{2}(\ell)$-based modifications to the DCF method \citep{houde2009,houde2016}
%Towards the Musca region, where the \bperp\ dispersion and the effect of the ``jitter'' are small, this would reduce by a factor of two the magnetic field strength estimate derived from the \Planck\ 353\,GHz observations \citep{planck2015-XXXV}.
%However, towards the other regions, where the ``jitter'' is larger, these differences introduced by the \bperp\ structure would amount to less than ten percent.

\section{Conclusions}\label{section:conclusions}

In this work we compared the orientation of the interstellar magnetic field averaged along the LOS and projected on the plane of the sky inferred from starlight and submillimetre polarization towards four nearby MCs. 
We found that the difference in the field orientation sampled in emission and extinction is on average less than 5\degr, thus suggesting considerable agreement between the \bperp\ orientations estimated with both techniques at comparable scales.

We evaluated the dispersion of the \bperp\ orientation, $\varsigma_{\psi_{\rm star}}$, inferred from starlight polarization within regions with the same area as the 353\,GHz \planck\ effective beam. 
We found that the values of $\varsigma_{\psi_{\rm star}}$ are on average less than 20\deg.
This result is in rough agreement with the values found using the angular structure function, $S_{2}(\ell)$, at scales below the size of the \planck\ beam $\ell < 10$\arcmin.
Given the current lack of information on the distance to the majority of the stars in the present work, the conclusions on the magnetic field morphology in and beyond the cloud remains open. 
However, we will deepen the study of the relation between $p_{\rm star}$ and $P_{\rm submm}$ and the effect of dust grain alignment in a separate publication (Alves et al. 2016, in preparation).

We found that a Gaussian model of the turbulent structure of the magnetic field, introduced in Sect.~\ref{section:GaussianRealization},  reproduces the values of $\varsigma_{\psi_{\rm star}}$ and $\varsigma_{p_{\rm star}}/\bar{p}_{\rm star}$ towards the observed regions.
The correlation between these two quantities, shown in Fig.~\ref{fig:SigmapOverMeanpVsSigmaPsi}, is not possible to reproduce in terms of the dust grain alignment efficiency alone.
This extends the results of \cite{planck2014-XX} to the high angular resolution accessible through the starlight polarization observations, suggesting that $\varsigma_{\psi_{\rm star}}$ and $\varsigma_{p_{\rm star}}/\bar{p}_{\rm star}$ at these gas column densities, \nh\,$< 10^{22}$\,cm$^{-2}$, is mainly produced by fluctuations in the magnetic field structure, rather than to changes in grain shape and/or the efficiency of grain alignment.

In terms of \stwo, we also found a significant amount of structure at scales lower than the size of the \Planck\ beam, represented in differences $\Delta S_{2}(\ell)$ up to 20\pdeg7.
This structure is also reproduced by the Gaussian model and can be explained in terms of the magnetic field structure and the difference in angular resolution between both types of observations.

At scales larger than the size of the \Planck\ beam, $\ell > 10$\arcmin, we found very good agreement between \stwostar\ and \stwosubmm. 
However, we find that the number and distribution of starlight polarization observations introduce oscillatory features, ``jitter'', in $S_{2}(\ell)$, which are not present when \stwo\ is calculated with a larger number of randomly-distributed observations, as it is now possible with the \planck\ observations.

The \stwo\ analysis of the Gaussian polarization models indicates that the differences between \stwostar\ and \stwosubmm\ depend on both the difference in the angular resolutions between the two types of polarization observation and the structure of the magnetic field.
When the field has a low amount of spatial correlations; as it is the case if the power spectrum of the turbulent field is relatively flat, or the turbulent field is relatively large with respect to the mean field, or the inclination of the mean with respect to the plane of the sky is relatively large; the differences between the observations at different angular resolutions can be large, even if the two techniques are sampling the same field.

Despite the presence of the ``jitter'' and the differences between \stwosubmm\  and \stwoallsubmm\ towards some of the regions, this study indicates that the increase in angular resolution, which is possible with the starlight polarization observations, indicates that the field structure in scales below that of the \Planck\ beam would not introduce significant corrections to the MC-scale magnetic field strengths estimated in \cite{planck2015-XXXV}.

We started this study looking to identify the effect of the angular resolution of the \Planck\ beam and constrain the portion of the line of sight that is responsible for the field orientations inferred from the \Planck\ 353\,GHz observations. 
We found magnetic field structure at scales below size of the \Planck\ beam and considerable agreement between both techniques at scales where they are comparable.
The question of the line of sight depth remains open, but soon, the advent of the catalog of distance observations by ESA's \emph{GAIA} satellite \citep{lindegren2010} will enable the study of the magnetic field morphology in different segments of the line of sight by correlating the distance to the stars to their polarization orientation and morphology. 

\begin{acknowledgements}
We thank M. Houde for his helpful comments.
This was possible through the funding from the European Research Council under the European
Community's Seventh Framework Programme (FP7/2007-2013 Grant Agreement no. 306483 and no. 291294).
\end{acknowledgements}

\bibliographystyle{aa}
\bibliography{Star25MAY2016.bbl}

\appendix

% ---------------------------------------------------------------------------------------------------------------------------------------------------------------------------------------------------------
%\begin{figure}[ht!]
%\centerline{
%\hspace{-0.5cm}
%\includegraphics[width=0.5\textwidth,angle=0,origin=c]{AllStarSubmm_MeanP353VsDiff.eps}
%}
%\caption{Scatter plot of the mean polarized flux in the \Planck\ 353\,GHz observations, $\bar{P}_{\rm submm}$, against the dispersion of orientation angles, $\varsigma_{\psi_{\rm star}}$ (\emph{left}), the fluctuations in polarized fraction, $\varsigma_{p_{\rm star}}/\bar{p}_{\rm star}$ (\emph{middle}), and orientation angle difference $\langle\bar{\psi}_{\rm star}-\bar{\psi}_{\rm submm}\rangle$ (\emph{right}), in all the 10\arcmin\ vicinities with more than three stars in each region.}
%\label{fig:MeanP353VsSigmapOverMeanp}
%\end{figure}

% ---------------------------------------------------------------------------------------------------------------------------------------------------------------------------------------------------------
\section{Histogram of relative orientations}\label{section:HRO}

In the main part of this work, we showed how the \bperp\ orientations inferred from starlight polarization observations follow closely those inferred from the \Planck\ observations when compared at the scale of 10\arcmin, within approximately 5\deg.
In this appendix, we evaluate if the \bperpstars\ observations can also be used to recover the trends found in the relative orientation of the magnetic field with respect to the column density structures reported in \cite{planck2015-XXXV}.

For that purpose we use the histogram of relative orientations (HRO) technique introduced in  \cite{soler2013} and \cite{planck2015-XXXV}, where the orientations of the column density structures are characterized by their gradients, which are by definition perpendicular to the iso-column density contours.
Here as in \cite{planck2015-XXXV} we use $\tau_{353}$ as a proxy for \nh\ (Sect.~\ref{data:columndensity}). 
The angle $\theta$ 
between \bperp\ and the tangent to the $\tau_{353}$ contours is evaluated using
\begin{equation}\label{eq:hroangle}
\theta = \arctan\left(\,|\nabla\,\tau_{353}\times \vec{\hat{E}}\,| \, , \, \nabla \tau_{353}\cdot \vec{\hat{E}}\,\right)\, ,
\end{equation}
where $\nabla\,\tau_{353}$ is perpendicular to the tangent of the iso-$\tau_{353}$ contours, the orientation of the unit polarization pseudo-vector $\vec{\hat{E}}$, perpendicular to \bperp, is characterized by the polarization angle $\theta_{\rm submm}$. In Eq.~\eqref{eq:hroangle}, as implemented, the norm actually carries a sign when the range used for $\theta$ is between $-90$\deg\ and $90$\deg.

We directly evaluate the relative orientations using both starlight and submillimetre polarization towards the four considered MCs, although the number of observations and their dynamic range in \nh\ values clearly limits the direct comparison with the results of \cite{planck2015-XXXV}. 

\begin{figure}[ht!]
\centerline{
\includegraphics[width=0.45\textwidth,angle=0,origin=c]{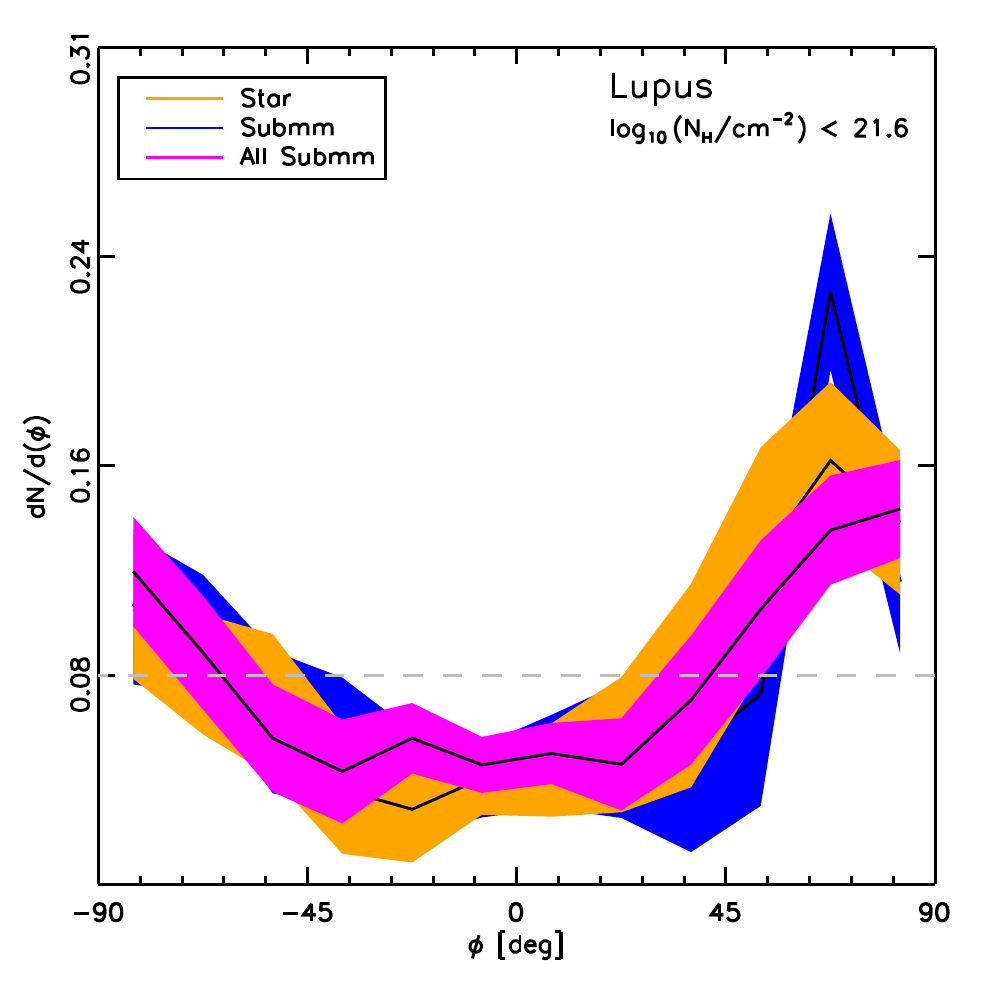}
}
\caption[]{Histograms of relative orientations between \bperp\ and the iso-$\tau_{353}$ contours towards Lupus. 
Histograms peaking at $90$\deg\ and/or $-90$\deg\ correspond to \bperp\ predominantly perpendicular to iso-$\tau_{353}$ contours.
In contrast with \cite{planck2015-XXXV}, the error bars are estimated by sampling of a Gaussian polarization model, including -by construction- spatial correlations which are absent when considering simple Poissonian statistics.}\label{fig:HROLupus}
\end{figure}

Fig.~\ref{fig:HROLupus} presents the HROs corresponding to LOSs with starlight polarization observations and \lognh\,$<21.6$ towards Lupus I, the region where the starlight polarization statistics were sufficient to produce a HRO consistent with what is found over the whole map with the \planck\ observations in the same column density regime. 
Despite the fact that previous studies have reported on relative orientation trends towards these regions using starlight polarization observations \citep{palmeirim2013,franco2015}, the systematic study of the change of relative orientation is for the moment only possible with the large statistics provided by the submillimetre polarization observations by \Planck.

\end{document}

%% file: Planck.tex
\def\setsymbol#1#2{\expandafter\def\csname #1\endcsname{#2}}
\def\getsymbol#1{\csname #1\endcsname}

\def\Planck{\textit{Planck}}

\def\all2013resultspapers{\nocite{planck2013-p01, planck2013-p02, planck2013-p02a, planck2013-p02d, planck2013-p02b, planck2013-p03, planck2013-p03c, planck2013-p03f, planck2013-p03d, planck2013-p03e, planck2013-p01a, planck2013-p06, planck2013-p03a, planck2013-pip88, planck2013-p08, planck2013-p11, planck2013-p12, planck2013-p13, planck2013-p14, planck2013-p15, planck2013-p05b, planck2013-p17, planck2013-p09, planck2013-p09a, planck2013-p20, planck2013-p19, planck2013-pipaberration, planck2013-p05, planck2013-p05a, planck2013-pip56, planck2013-p06b, planck2013-p01a}}

\newbox\tablebox    \newdimen\tablewidth
\def\leaderfil{\leaders\hbox to 5pt{\hss.\hss}\hfil}
\def\endPlancktablewide{\tablewidth=\textwidth 
    $$\hss\copy\tablebox\hss$$
    \vskip-\lastskip\vskip -2pt}
\def\tablenote#1 #2\par{\begingroup \parindent=0.8em
    \abovedisplayshortskip=0pt\belowdisplayshortskip=0pt
    \noindent
    $$\hss\vbox{\hsize\tablewidth \hangindent=\parindent \hangafter=1 \noindent
    \hbox to \parindent{$^#1$\hss}\strut#2\strut\par}\hss$$
    \endgroup}
\def\doubleline{\vskip 3pt\hrule \vskip 1.5pt \hrule \vskip 5pt}

\def\L2{\ifmmode L_2\else $L_2$\fi}

\def\DeltaT{\ifmmode \Delta T\else $\Delta T$\fi}
\def\deltat{\ifmmode \Delta t\else $\Delta t$\fi}
\def\fknee{\ifmmode f_{\rm knee}\else $f_{\rm knee}$\fi}
\def\Fmax{\ifmmode F_{\rm max}\else $F_{\rm max}$\fi}
\def\solar{\ifmmode{\rm M}_{\mathord\odot}\else${\rm M}_{\mathord\odot}$\fi}
\def\Msolar{\ifmmode{\rm M}_{\mathord\odot}\else${\rm M}_{\mathord\odot}$\fi}
\def\Lsolar{\ifmmode{\rm L}_{\mathord\odot}\else${\rm L}_{\mathord\odot}$\fi}
\def\inv{\ifmmode^{-1}\else$^{-1}$\fi}
\def\mo{\ifmmode^{-1}\else$^{-1}$\fi}
\def\sup#1{\ifmmode ^{\rm #1}\else $^{\rm #1}$\fi}
\def\expo#1{\ifmmode \times 10^{#1}\else $\times 10^{#1}$\fi}
\def\,{\thinspace}
\def\lsim{\mathrel{\raise .4ex\hbox{\rlap{$<$}\lower 1.2ex\hbox{$\sim$}}}}
\def\gsim{\mathrel{\raise .4ex\hbox{\rlap{$>$}\lower 1.2ex\hbox{$\sim$}}}}

\def\simprop{\mathrel{\raise .4ex\hbox{\rlap{$\propto$}\lower 1.2ex\hbox{$\sim$}}}}
\def\deg{\ifmmode^\circ\else$^\circ$\fi}
\def\pdeg{\ifmmode $\setbox0=\hbox{$^{\circ}$}\rlap{\hskip.11\wd0 .}$^{\circ}
          \else \setbox0=\hbox{$^{\circ}$}\rlap{\hskip.11\wd0 .}$^{\circ}$\fi}
\def\arcs{\ifmmode {^{\scriptstyle\prime\prime}}
          \else $^{\scriptstyle\prime\prime}$\fi}
\def\arcm{\ifmmode {^{\scriptstyle\prime}}
          \else $^{\scriptstyle\prime}$\fi}
\newdimen\sa  \newdimen\sb
\def\parcs{\sa=.07em \sb=.03em
     \ifmmode \hbox{\rlap{.}}^{\scriptstyle\prime\kern -\sb\prime}\hbox{\kern -\sa}
     \else \rlap{.}$^{\scriptstyle\prime\kern -\sb\prime}$\kern -\sa\fi}
\def\parcm{\sa=.08em \sb=.03em
     \ifmmode \hbox{\rlap{.}\kern\sa}^{\scriptstyle\prime}\hbox{\kern-\sb}
     \else \rlap{.}\kern\sa$^{\scriptstyle\prime}$\kern-\sb\fi}
\def\ra[#1 #2 #3.#4]{#1\sup{h}#2\sup{m}#3\sup{s}\llap.#4}
\def\dec[#1 #2 #3.#4]{#1\deg#2\arcm#3\arcs\llap.#4}
\def\deco[#1 #2 #3]{#1\deg#2\arcm#3\arcs}
\def\rra[#1 #2]{#1\sup{h}#2\sup{m}}

\def\dots{\relax\ifmmode \ldots\else $\ldots$\fi}
\def\WHzsr{\ifmmode $W\,Hz\mo\,sr\mo$\else W\,Hz\mo\,sr\mo\fi}
\def\mHz{\ifmmode $\,mHz$\else \,mHz\fi}
\def\GHz{\ifmmode $\,GHz$\else \,GHz\fi}
\def\mKs{\ifmmode $\,mK\,s$^{1/2}\else \,mK\,s$^{1/2}$\fi}
\def\muKs{\ifmmode \,\mu$K\,s$^{1/2}\else \,$\mu$K\,s$^{1/2}$\fi}
\def\muKRJs{\ifmmode \,\mu$K$_{\rm RJ}$\,s$^{1/2}\else \,$\mu$K$_{\rm RJ}$\,s$^{1/2}$\fi}
\def\muKHz{\ifmmode \,\mu$K\,Hz$^{-1/2}\else \,$\mu$K\,Hz$^{-1/2}$\fi}
\def\MJysr{\ifmmode \,$MJy\,sr\mo$\else \,MJy\,sr\mo\fi}
\def\MJysrmK{\ifmmode \,$MJy\,sr\mo$\,mK$_{\rm CMB}\mo\else \,MJy\,sr\mo\,mK$_{\rm CMB}\mo$\fi}
\def\microns{\ifmmode \,\mu$m$\else \,$\mu$m\fi}
\def\micron{\microns}
\def\muK{\ifmmode \,\mu$K$\else \,$\mu$\hbox{K}\fi}
\def\microK{\ifmmode \,\mu$K$\else \,$\mu$\hbox{K}\fi}
\def\muW{\ifmmode \,\mu$W$\else \,$\mu$\hbox{W}\fi}
\def\kms{\ifmmode $\,km\,s$^{-1}\else \,km\,s$^{-1}$\fi}
\def\kmsMpc{\ifmmode $\,\kms\,Mpc\mo$\else \,\kms\,Mpc\mo\fi}
\providecommand{\sorthelp}[1]{}

%% file: Star25MAY2016.bbl
\begin{thebibliography}{53}
\expandafter\ifx\csname natexlab\endcsname\relax\def\natexlab#1{#1}\fi

\bibitem[{{Alves} \& {Franco}(2007)}]{alves2007}
{Alves}, F.~O. \& {Franco}, G.~A.~P. 2007, \aap, 470, 597

\bibitem[{{Andersson} {et~al.}(2015){Andersson}, {Lazarian}, \&
  {Vaillancourt}}]{andersson2015}
{Andersson}, B.-G., {Lazarian}, A., \& {Vaillancourt}, J.~E. 2015, \araa, 53,
  501

\bibitem[{{Bergin} \& {Tafalla}(2007)}]{bergin2007}
{Bergin}, E.~A. \& {Tafalla}, M. 2007, \araa, 45, 339

\bibitem[{{Cabral} \& {Leedom}(1993)}]{cabral1993}
{Cabral}, B. \& {Leedom}, L.~C. 1993, in Special Interest Group on GRAPHics and
  Interactive Techniques Proceedings., Special Interest Group on GRAPHics and
  Interactive Techniques Proceedings.

\bibitem[{{Chandrasekhar} \& {Fermi}(1953)}]{chandrasekhar1953}
{Chandrasekhar}, S. \& {Fermi}, E. 1953, \apj, 118, 113

\bibitem[{{Chapman} {et~al.}(2011){Chapman}, {Goldsmith}, {Pineda}, {Clemens},
  {Li}, \& {Kr{\v c}o}}]{chapman2011}
{Chapman}, N.~L., {Goldsmith}, P.~F., {Pineda}, J.~L., {et~al.} 2011, \apj,
  741, 21

\bibitem[{{Clemens} {et~al.}(2007){Clemens}, {Sarcia}, {Grabau}, {Tollestrup},
  {Buie}, {Dunham}, \& {Taylor}}]{clemens2007}
{Clemens}, D.~P., {Sarcia}, D., {Grabau}, A., {et~al.} 2007, \pasp, 119, 1385

\bibitem[{{Crutcher}(2012)}]{crutcher2012}
{Crutcher}, R.~M. 2012, \araa, 50, 29

\bibitem[{{Dame} {et~al.}(2001){Dame}, {Hartmann}, \& {Thaddeus}}]{dame2001}
{Dame}, T.~M., {Hartmann}, D., \& {Thaddeus}, P. 2001, \apj, 547, 792

\bibitem[{{Davis}(1951)}]{davis1951a}
{Davis}, L. 1951, Physical Review, 81, 890

\bibitem[{{Davis} \& {Greenstein}(1951)}]{davis1951}
{Davis}, Jr., L. \& {Greenstein}, J.~L. 1951, \apj, 114, 206

\bibitem[{{Falceta-Gon{\c c}alves} {et~al.}(2008){Falceta-Gon{\c c}alves},
  {Lazarian}, \& {Kowal}}]{falceta2008}
{Falceta-Gon{\c c}alves}, D., {Lazarian}, A., \& {Kowal}, G. 2008, \apj, 679,
  537

\bibitem[{{Franco} \& {Alves}(2015)}]{franco2015}
{Franco}, G.~A.~P. \& {Alves}, F.~O. 2015, \apj, 807, 5

\bibitem[{{Franco} {et~al.}(2010){Franco}, {Alves}, \& {Girart}}]{franco2010}
{Franco}, G.~A.~P., {Alves}, F.~O., \& {Girart}, J.~M. 2010, \apj, 723, 146

\bibitem[{{Gaczkowski} {et~al.}(2015){Gaczkowski}, {Preibisch}, {Stanke},
  {Krause}, {Burkert}, {Diehl}, {Fierlinger}, {Kroell}, {Ngoumou}, \&
  {Roccatagliata}}]{gaczkowski2015}
{Gaczkowski}, B., {Preibisch}, T., {Stanke}, T., {et~al.} 2015, \aap, 584, A36

\bibitem[{{G{\'o}rski} {et~al.}(2005){G{\'o}rski}, {Hivon}, {Banday},
  {Wandelt}, {Hansen}, {Reinecke}, \& {Bartelmann}}]{gorski2005}
{G{\'o}rski}, K.~M., {Hivon}, E., {Banday}, A.~J., {et~al.} 2005, \apj, 622,
  759

\bibitem[{{Hamaker} \& {Bregman}(1996)}]{hamaker1996III}
{Hamaker}, J.~P. \& {Bregman}, J.~D. 1996, \aaps, 117, 161

\bibitem[{{Heiles}(2000)}]{heiles2000}
{Heiles}, C. 2000, \aj, 119, 923

\bibitem[{{Hildebrand}(1988)}]{hildebrand1988}
{Hildebrand}, R.~H. 1988, \qjras, 29, 327

\bibitem[{{Hildebrand} {et~al.}(2009){Hildebrand}, {Kirby}, {Dotson}, {Houde},
  \& {Vaillancourt}}]{hildebrand2009}
{Hildebrand}, R.~H., {Kirby}, L., {Dotson}, J.~L., {Houde}, M., \&
  {Vaillancourt}, J.~E. 2009, \apj, 696, 567

\bibitem[{{Hiltner}(1949)}]{hiltner1949}
{Hiltner}, W.~A. 1949, Science, 109, 165

\bibitem[{{Houde} {et~al.}(2016){Houde}, {Hull}, {Plambeck}, {Vaillancourt}, \&
  {Hildebrand}}]{houde2016}
{Houde}, M., {Hull}, C.~L.~H., {Plambeck}, R.~L., {Vaillancourt}, J.~E., \&
  {Hildebrand}, R.~H. 2016, \apj, 820, 38

\bibitem[{{Houde} {et~al.}(2009){Houde}, {Vaillancourt}, {Hildebrand},
  {Chitsazzadeh}, \& {Kirby}}]{houde2009}
{Houde}, M., {Vaillancourt}, J.~E., {Hildebrand}, R.~H., {Chitsazzadeh}, S., \&
  {Kirby}, L. 2009, \apj, 706, 1504

\bibitem[{{Kenyon} {et~al.}(2008){Kenyon}, {G{\'o}mez}, \&
  {Whitney}}]{kenyon2008}
{Kenyon}, S.~J., {G{\'o}mez}, M., \& {Whitney}, B.~A. 2008, {Low Mass Star
  Formation in the Taurus-Auriga Clouds}, ed. B.~{Reipurth}, 405

\bibitem[{{Kobulnicky} {et~al.}(1994){Kobulnicky}, {Molnar}, \&
  {Jones}}]{kobulnicky1994}
{Kobulnicky}, H.~A., {Molnar}, L.~A., \& {Jones}, T.~J. 1994, \aj, 107, 1433

\bibitem[{{Lamarre} {et~al.}(2010){Lamarre}, {Puget}, {Ade}, {Bouchet},
  {Guyot}, {Lange}, {Pajot}, {Arondel}, {Benabed}, {Beney}, {Beno{\^i}t},
  {Bernard}, {Bhatia}, {Blanc}, {Bock}, {Br{\'e}elle}, {Bradshaw}, {Camus},
  {Catalano}, {Charra}, {Charra}, {Church}, {Couchot}, {Coulais}, {Crill},
  {Crook}, {Dassas}, {de Bernardis}, {Delabrouille}, {de Marcillac}, {Delouis},
  {D{\'e}sert}, {Dumesnil}, {Dupac}, {Efstathiou}, {Eng}, {Evesque},
  {Fourmond}, {Ganga}, {Giard}, {Gispert}, {Guglielmi}, {Haissinski},
  {Henrot-Versill{\'e}}, {Hivon}, {Holmes}, {Jones}, {Koch}, {Lagard{\`e}re},
  {Lami}, {Land{\'e}}, {Leriche}, {Leroy}, {Longval},
  {Mac{\'{\i}}as-P{\'e}rez}, {Maciaszek}, {Maffei}, {Mansoux}, {Marty}, {Masi},
  {Mercier}, {Miville-Desch{\^e}nes}, {Moneti}, {Montier}, {Murphy},
  {Narbonne}, {Nexon}, {Paine}, {Pahn}, {Perdereau}, {Piacentini}, {Piat},
  {Plaszczynski}, {Pointecouteau}, {Pons}, {Ponthieu}, {Prunet}, {Rambaud},
  {Recouvreur}, {Renault}, {Ristorcelli}, {Rosset}, {Santos}, {Savini},
  {Serra}, {Stassi}, {Sudiwala}, {Sygnet}, {Tauber}, {Torre}, {Tristram},
  {Vibert}, {Woodcraft}, {Yurchenko}, \& {Yvon}}]{lamarre2010}
{Lamarre}, J., {Puget}, J., {Ade}, P.~A.~R., {et~al.} 2010, \aap, 520, A9

\bibitem[{{Lazarian} \& {Hoang}(2007)}]{lazarian2007}
{Lazarian}, A. \& {Hoang}, T. 2007, \mnras, 378, 910

\bibitem[{{Lindegren}(2010)}]{lindegren2010}
{Lindegren}, L. 2010, in IAU Symposium, Vol. 261, Relativity in Fundamental
  Astronomy: Dynamics, Reference Frames, and Data Analysis, ed. S.~A.
  {Klioner}, P.~K. {Seidelmann}, \& M.~H. {Soffel}, 296--305

\bibitem[{{Luhman}(2008)}]{luhman2008}
{Luhman}, K.~L. 2008, {Chamaeleon}, ed. B.~{Reipurth}, 169

\bibitem[{{Magalhaes} {et~al.}(1996){Magalhaes}, {Rodrigues}, {Margoniner},
  {Pereyra}, \& {Heathcote}}]{magalhaes1996}
{Magalhaes}, A.~M., {Rodrigues}, C.~V., {Margoniner}, V.~E., {Pereyra}, A., \&
  {Heathcote}, S. 1996, in Astronomical Society of the Pacific Conference
  Series, Vol.~97, Polarimetry of the Interstellar Medium, ed. W.~G. {Roberge}
  \& D.~C.~B. {Whittet}, 118

\bibitem[{{Martin} {et~al.}(2012){Martin}, {Roy}, {Bontemps},
  {Miville-Desch{\^e}nes}, {Ade}, {Bock}, {Chapin}, {Devlin}, {Dicker},
  {Griffin}, {Gundersen}, {Halpern}, {Hargrave}, {Hughes}, {Klein}, {Marsden},
  {Mauskopf}, {Netterfield}, {Olmi}, {Patanchon}, {Rex}, {Scott}, {Semisch},
  {Truch}, {Tucker}, {Tucker}, {Viero}, \& {Wiebe}}]{martin2012}
{Martin}, P.~G., {Roy}, A., {Bontemps}, S., {et~al.} 2012, \apj, 751, 28

\bibitem[{{McKee} \& {Ostriker}(2007)}]{mckee2007}
{McKee}, C.~F. \& {Ostriker}, E.~C. 2007, \araa, 45, 565

\bibitem[{{Montier} {et~al.}(2015){Montier}, {Plaszczynski}, {Levrier},
  {Tristram}, {Alina}, {Ristorcelli}, {Bernard}, \& {Guillet}}]{montier2015}
{Montier}, L., {Plaszczynski}, S., {Levrier}, F., {et~al.} 2015, \aap, 574,
  A136

\bibitem[{{Naghizadeh-Khouei} \& {Clarke}(1993)}]{naghizadeh-khouei1993}
{Naghizadeh-Khouei}, J. \& {Clarke}, D. 1993, \aap, 274, 968

\bibitem[{{Palmeirim} {et~al.}(2013){Palmeirim}, {Andr{\'e}}, {Kirk},
  {Ward-Thompson}, {Arzoumanian}, {K{\"o}nyves}, {Didelon}, {Schneider},
  {Benedettini}, {Bontemps}, {Di Francesco}, {Elia}, {Griffin}, {Hennemann},
  {Hill}, {Martin}, {Men'shchikov}, {Molinari}, {Motte}, {Nguyen Luong},
  {Nutter}, {Peretto}, {Pezzuto}, {Roy}, {Rygl}, {Spinoglio}, \&
  {White}}]{palmeirim2013}
{Palmeirim}, P., {Andr{\'e}}, P., {Kirk}, J., {et~al.} 2013, \aap, 550, A38

\bibitem[{{Paradis} {et~al.}(2012){Paradis}, {Dobashi}, {Shimoikura},
  {Kawamura}, {Onishi}, {Fukui}, \& {Bernard}}]{paradis2012}
{Paradis}, D., {Dobashi}, K., {Shimoikura}, T., {et~al.} 2012, \aap, 543, A103

\bibitem[{{Pereyra} \& {Magalh{\~a}es}(2004)}]{pereyra2004}
{Pereyra}, A. \& {Magalh{\~a}es}, A.~M. 2004, \apj, 603, 584

\bibitem[{{\sorthelp{Planck Collaboration 2011X}}{Planck Collaboration
  XXIV}(2011)}]{planck2011-7.12}
{\sorthelp{Planck Collaboration 2011X}}{Planck Collaboration XXIV}. 2011, \aap,
  536, A24

\bibitem[{{\sorthelp{Planck Collaboration 2014A}}{Planck Collaboration
  I}(2014)}]{planck2013-p01}
{\sorthelp{Planck Collaboration 2014A}}{Planck Collaboration I}. 2014, \aap,
  571, A1

\bibitem[{{\sorthelp{Planck Collaboration 2014K}}{Planck Collaboration
  XI}(2014)}]{planck2013-p06b}
{\sorthelp{Planck Collaboration 2014K}}{Planck Collaboration XI}. 2014, \aap,
  571, A11

\bibitem[{{\sorthelp{Planck Collaboration 2015A}}{Planck Collaboration
  I}(2016)}]{planck2014-a01}
{\sorthelp{Planck Collaboration 2015A}}{Planck Collaboration I}. 2016, \aap,
  submitted

\bibitem[{{\sorthelp{Planck Collaboration IntS}}{Planck Collaboration Int.
  XIX}(2015)}]{planck2014-XIX}
{\sorthelp{Planck Collaboration IntS}}{Planck Collaboration Int. XIX}. 2015,
  \aap, 576, A104

\bibitem[{{\sorthelp{Planck Collaboration IntT}}{Planck Collaboration Int.
  XX}(2015)}]{planck2014-XX}
{\sorthelp{Planck Collaboration IntT}}{Planck Collaboration Int. XX}. 2015,
  \aap, 576, A105

\bibitem[{{\sorthelp{Planck Collaboration IntU}}{Planck Collaboration Int.
  XXI}(2015)}]{planck2014-XXI}
{\sorthelp{Planck Collaboration IntU}}{Planck Collaboration Int. XXI}. 2015,
  \aap, 576, A106

\bibitem[{{\sorthelp{Planck Collaboration IntZD}}{Planck Collaboration Int.
  XXIX}(2016)}]{planck2014-XXIX}
{\sorthelp{Planck Collaboration IntZD}}{Planck Collaboration Int. XXIX}. 2016,
  \aap, 586, A132

\bibitem[{{\sorthelp{Planck Collaboration IntZG}}{Planck Collaboration Int.
  XXXII}(2016)}]{planck2014-XXXII}
{\sorthelp{Planck Collaboration IntZG}}{Planck Collaboration Int. XXXII}. 2016,
  \aap, 586, A135

\bibitem[{{\sorthelp{Planck Collaboration IntZH}}{Planck Collaboration Int.
  XXXIII}(2016)}]{planck2014-XXXIII}
{\sorthelp{Planck Collaboration IntZH}}{Planck Collaboration Int. XXXIII}.
  2016, \aap, 586, A136

\bibitem[{{\sorthelp{Planck Collaboration IntZJ}}{Planck Collaboration Int.
  XXXV}(2016)}]{planck2015-XXXV}
{\sorthelp{Planck Collaboration IntZJ}}{Planck Collaboration Int. XXXV}. 2016,
  \aap, 586, A138

\bibitem[{{\sorthelp{Planck Collaboration IntZT}}{Planck Collaboration Int.
  XLV}(2016)}]{planck2016-XLV}
{\sorthelp{Planck Collaboration IntZT}}{Planck Collaboration Int. XLV}. 2016,
  in preparation

\bibitem[{{Rygl} {et~al.}(2013){Rygl}, {Benedettini}, {Schisano}, {Elia},
  {Molinari}, {Pezzuto}, {Andr{\'e}}, {Bernard}, {White}, {Polychroni},
  {Bontemps}, {Cox}, {Di Francesco}, {Facchini}, {Fallscheer}, {di Giorgio},
  {Hennemann}, {Hill}, {K{\"o}nyves}, {Minier}, {Motte}, {Nguyen-Luong},
  {Peretto}, {Pestalozzi}, {Sadavoy}, {Schneider}, {Spinoglio}, {Testi}, \&
  {Ward-Thompson}}]{rygl2013}
{Rygl}, K.~L.~J., {Benedettini}, M., {Schisano}, E., {et~al.} 2013, \aap, 549,
  L1

\bibitem[{{Schlafly} {et~al.}(2014){Schlafly}, {Green}, {Finkbeiner}, {Rix},
  {Bell}, {Burgett}, {Chambers}, {Draper}, {Hodapp}, {Kaiser}, {Magnier},
  {Martin}, {Metcalfe}, {Price}, \& {Tonry}}]{schlafly2014}
{Schlafly}, E.~F., {Green}, G., {Finkbeiner}, D.~P., {et~al.} 2014, \apj, 786,
  29

\bibitem[{{Serkowski}(1958)}]{serkowski1958}
{Serkowski}, K. 1958, \actaa, 8, 135

\bibitem[{{Soler} {et~al.}(2013){Soler}, {Hennebelle}, {Martin},
  {Miville-Desch{\^e}nes}, {Netterfield}, \& {Fissel}}]{soler2013}
{Soler}, J.~D., {Hennebelle}, P., {Martin}, P.~G., {et~al.} 2013, \apj, 774,
  128

\end{thebibliography}
